\documentstyle[aps,preprint,floats,epsfig]{revtex}
\begin{document}
\tighten
\title{An Analysis of the Next-to-Leading Order\\
Corrections to the $g_T(=g_1+g_2)$ Scaling Function}

\author{Xiangdong Ji$^*$ and Jonathan Osborne$^\dagger$}
\address{$^*$Department of Physics, University of Maryland\\
College Park, MD 20742}
\address{$^\dagger$Nuclear Science Division, Lawrence Berkeley 
National Laboratory\\
Berkeley, CA 94720}

\date{UMD PP\#01-035, LBNL-47474 ~~~DOE/ER/40762-219~~~ Feburary 2001}
\maketitle

\begin{abstract}

We present a general method for obtaining the 
quantum chromodynamical radiative corrections 
to the 
higher-twist (power-suppressed) 
contributions to inclusive deep-inelastic scattering
in terms of light-cone correlation functions
of the fundamental fields of quantum chromodynamics.
Using this procedure, 
we calculate the previously unknown ${\cal O}(\alpha_s)$
corrections to the twist-three part of the 
spin scaling function $g_T(x_B,Q^2)
(=g_1(x_B,Q^2)+g_2(x_B,Q^2))$ and the corresponding 
forward Compton amplitude $S_T(\nu,Q^2)$.
Expanding our result about the unphysical point $x_B=\infty$,
we arrive at an operator product expansion of the 
nonlocal product of two electromagnetic current operators 
involving twist-two and -three operators valid to ${\cal O}(\alpha_s)$
for forward matrix elements.
We find that the Wandzura-Wilczek relation between
$g_1(x_B,Q^2)$ and the twist-two part of 
$g_T(x_B,Q^2)$ is respected in both the 
singlet and non-singlet sectors 
at this order, and argue its validity to all orders.  
The large-$N_c$ limit does not appreciably
simplify the twist-three Wilson coefficients.

\end{abstract}
\pacs{xxxxxxxx}

\narrowtext

\section{Introduction}

Deep-inelastic scattering (DIS) of leptons on the nucleon 
is a time-honored example of the success
of perturbative quantum chromodynamics (PQCD)\cite{mueller}. 
The factorization formulae for the scaling  
functions 
$F_1(x_B, Q^2)$ and $g_1(x_B, Q^2)$
(defined from the Bjorken limit of the 
structure functions $W_1(\nu, Q^2)$ and $G_1(\nu, Q^2)$,
respectively), 
augmented by the Dokshitzer-
Gribov-Lipatov-Altarelli-Parisi (DGLAP) 
evolution equations for parton distributions
\cite{dglap}, describe the 
available DIS data collected over the last 
30 years exceedingly 
well \cite{cteq}. Although the same formalism is believed
to work for the so-called higher-twist structure 
functions \cite{sterman}, e.g. $W_L(\nu, Q^2)$ and $G_2(\nu, Q^2)$,
and their associated scaling functions,
$F_L(x_B,Q^2)$ and $g_2(x_B,Q^2)$,
which contribute
to physical observables down by powers of the hard 
momentum $Q$, there are few detailed 
studies of them in the literature beyond the tree level. 
[Actually, $F_L(x_B,Q^2)$ receives an $\alpha_s$-suppressed
twist-two contribution which we will ignore in this paper.]
The QCD radiative corrections to 
$F_L(x_B,Q^2)$ and $g_2(x_B,Q^2)$ need be investigated as accurate
data have recently been taken \cite{fldata,e155x} and more data will 
be available in the future \cite{jlab}.

In this paper, we present a general method for obtaining
radiative corrections to the higher-twist parts of structure
functions in inclusive DIS.  
This method is based on a generalization of the 
tree-level formalism 
for higher-twist structure functions presented
in Ref. \cite{cfp}, and has been outlined in 
a short paper of ours published earlier \cite{ournonsing}. 
From this, we extract a 
straightforward set of rules which can be used to calculate 
the amplitude of inclusive DIS at an arbitrary order in 
$\alpha_s$ and $\Lambda_{\rm QCD}/Q$, 
where $\Lambda_{\rm QCD}$ is the QCD scale, in terms
of light-cone correlation functions of quarks and gluons.
As an illustration of our method, we 
present a detailed calculation 
of the radiative corrections to the lowest higher-twist
observable, $G_2(\nu,Q^2)$.  The partial results of this 
calculation have been communicated earlier 
in Refs. \cite{ournonsing,oursing}.

In inclusive polarized deep-inelastic scattering
off a nucleon target, one can measure the spin-dependent
structure functions $G_{1,2}(\nu, Q^2)$. $G_1(\nu, Q^2)$
is closely related to the spin structure of the nucleon,
and has been studied extensively, both experimentally and 
theoretically, in the last few decades\cite{g1summary}.  
For quite a long period, the physics associated with 
$G_2(\nu,Q^2)$, in particular the scattering mechanism 
and its role in understanding the internal structure of 
the nucleon, has generated much debate in the literature 
\cite{debate}. In earlier investigations, attempts were made 
to interpret the structure function using the naive 
parton model and its various extensions \cite{naive}.  
However, further studies showed that those results 
are incompatible with the factorization theorems of 
QCD \cite{nonaive}. Indeed, a QCD investigation 
shows that the scaling function $g_2(x_B,Q^2)$ 
arises from the effects of parton 
transverse momentum and coherent parton scattering. 
The new information about the nucleon structure 
contained in $G_2(\nu, Q^2)$ is found in its 
twist-three part \cite{twist3}, which reflects the 
quark-gluon correlations in the nucleon.
In fact, this structure function represents the 
simplest manifestation of these correlations
found in experiment.  In addition, much of the 
information contained in 
$G_2(\nu,Q^2)$ has phenomenological interest.  For
example, the third moment of its scaling function, 
$g_2(x_B,Q^2)$, is related to the response
of the chromo-electric and -magnetic fields 
inside the nucleon to its spin polarization \cite{response}.
This information can help us to understand the role
of the gluon fields in the nucleon, to test lattice
QCD calculations, and to construct more accurate models 
of nucleon structure.

It is useful at this point to remind the reader of
a fey key developments in studying the $G_2$
structure function. Before QCD was invented,
Burkhardt and Cottingham derived the super-convergence
sum rule $\int d\nu\, G_2(\nu, Q^2)=0$ 
based on a dispersion relation and an assumption about 
asymptotic properties of the related Compton 
amplitude \cite{bcrule}. In the context of QCD, 
a number of issues about $G_2(\nu,Q^2)$ have been 
clarified in recent years.
At tree level, $g_2(x_B,Q^2)$ is related to 
a seemingly simple quark distribution, $q_T(x,Q^2)$,
within the nucleon \cite{tree}. However, as first pointed out by 
Shuryak and Vainshtein \cite{mixing},
its leading-log evolution 
reveals this simple form to be deceptive.
In reality, one requires the introduction of 
a more general distribution to obtain a closed
evolution equation \cite{bkl}.  
As found by Ali, Braun, and Hiller \cite{abh},
a somewhat surprising simplification occurs
in the limit of large number of colors.
In this limit, the leading order evolution
of $q_T(x,Q^2)$ becomes autonomous in the 
nonsinglet sector. However,
it was later argued that this
simplification does not occur at higher orders
or in the singlet sector \cite{ourlargen}.
An analysis of the local operators associated
with the tree-level expressions for the moments
of $g_2(x_B,Q^2)$ reveals a relation between 
its twist-two part and the scaling function
$g_1(x_B,Q^2)$ \cite{wwrel}.  This relation 
makes it possible to extract the twist-three
part of $g_2(x_B,Q^2)$ from the data. Experimental
measurements of $G_2$ have been made by a number of
collaborations \cite{g2exp}. In addition, model 
calculations and lattice QCD yield interesting
insights of the twist-three part of the 
structure function \cite{modeletc}.

This paper is organized as follows.
In Section 2, a general discussion of the 
kinematics and structure functions associated with 
spin-dependent inclusive DIS is made.  
This section serves to familiarize the reader with
some of our conventions and notations.
Section 3 is broken into five subsections 
dedicated to a systematic presentation
of the method we use to obtain radiative corrections
to the power-suppressed contributions to 
our amplitude.  First, we summarize the well-known 
light-cone power counting techniques to the process
at hand, which allows us to organize contributions to the 
amplitude with respect to their power suppression.
Then, we discuss the role of longitudinal gluons in this
process and the simplifications associated with the 
light-cone gauge.  This is followed by a simple example in which 
the known tree-level result for $G_2(\nu,Q^2)$ is derived.
It is here that the nonsinglet distributions in
inclusive DIS with transverse polarization are introduced.
Subsection D discusses the subtleties of renormalization and 
factorization relevant to radiative corrections.  The last
subsection presents a set of rules derived from this method which
can be used to obtain the contribution to inclusive DIS 
at an arbitrary order in $1/Q$ and $\alpha_s(Q^2)$.
In Section 4, we apply this method to 
the one-loop corrections of the results in 
Section 3C in the non-singlet and singlet sectors.  
These results were first presented 
in Refs. \cite{ournonsing,oursing}; in the 
present communication, we go into somewhat more
detail in explaining their derivation.
An earlier analysis of the 
one-loop corrections in the singlet sector can be found
in Ref.\cite{oldsing}. Although the preliminary 
amplitudes are the same, the final answer and 
its interpretation are quite different.
A recent study of these corrections \cite{shoot}
presents the same interpretation found here.
There, one can also find discussions of the numerical
significance of the results.
Section 5 presents the results of the last section in the 
form of an operator product expansion.  In this form, 
it is easy to exploit the full space-time 
symmetries of our theory and isolate the new information
provided by transverse scattering.  
This is one of the main results of this paper, 
as the twist-three Wilson coefficients 
have never been obtained at one-loop order before.
In Section 6, 
we explain how to perform the twist separation
without the use of the operator product expansion.
Section 7 contains some concluding remarks and 
future possible work.

\section{Kinematics}

To familiarize the reader with our conventions and 
notations, we review in this section a number of well-known
facts about polarized deep-inelastic scattering (DIS).
The process under consideration is 
inclusive lepton-nucleon scattering
through one photon exchange.  
Although we have electron-proton scattering in mind, our results
are applicable to processes involving any charged lepton and any
hadron.  
Writing out the scattering amplitude, squaring,
and summing over final states, one finds to leading order in 
$\alpha_{\rm em}$ that the cross-section factors into
two pieces :
\begin{eqnarray}
\sigma&\propto& {4\pi\alpha_{\rm em}\over q^4}
L^{\mu\nu}W_{\mu\nu}\;\; ; \nonumber\\
W^{\mu\nu}(q,P,S)&=
&{1\over4\pi}\int d^{\,4}z\,e^{iq\cdot z}\,\left\langle
PS\left|\left\lbrack J^{\mu}(z),J^\nu(0)\right\rbrack
\right|PS\right\rangle\;\; ,
\label{Wdef}
\end{eqnarray}
where we have taken $q^\mu$ as the virtual photon momentum.  $J^\mu$
is the electromagnetic current operator,
\begin{equation}
J^\mu(z)=\sum_qe_q\overline\psi_q(z)\gamma^\mu\psi_q(z)\;\; ,
\end{equation}
and $L^{\mu\nu}$ is a tensor that depends only on the 
lepton spin and momentum.  The ket $|PS\rangle$
represents a nucleon state 
of momentum $P^\mu$ and spin
polarization $S^\mu$, which are constrained by 
$P^2=M^2$, $S^2=-1$, and $P\cdot S=0$.
The nontrivial kinematical invariants are 
$Q^2\equiv-q^2$ and $\nu\equiv q\cdot P$, which are
positive in the physical domain.

In polarized scattering, only the $\mu\nu$ antisymmetric 
part of $W^{\mu\nu}$ contributes.  Using discrete 
spacetime symmetries, one can 
express it in terms of two independent structure functions : 
\begin{equation}
W^{[\mu\nu]}=-i\epsilon^{\mu\nu\alpha\beta}q_\alpha 
{1\over M}\left\lbrack S_\beta \,G_1(\nu,Q^2)
+{1\over M^2}\,\left(\nu\; S-q\cdot S\; P
\right)_\beta\,G_2(\nu,Q^2)\right\rbrack\;\; ,
\label{decomp}
\end{equation}
where $[\cdots]$ denotes antisymmetrization
and $\epsilon^{\mu\nu\alpha\beta}$ is the Levi-Cevita tensor
($\epsilon^{0123}=+1$).  All of the structural information
on our nucleon state is contained in the dimensionless functions
$G_{1,2}(\nu,Q^2)$.  Through the optical theorem, one can relate
these structure functions to the invariant 
amplitudes contained in the forward Compton amplitude,
\begin{equation}
T^{\mu\nu}=i\int d^{\,4}z\,e^{iq\cdot z}\left\langle PS\left|
{\cal T}\left\lbrace J^\mu(z)
J^\nu(0)\right\rbrace\right|PS\right\rangle\;\; ,
\label{Tdef}
\end{equation}
as
\begin{eqnarray}
\label{sdecomp}
&&T^{[\mu\nu]}=-i\epsilon^{\mu\nu\alpha\beta}
q_\alpha\,{1\over M}\left\lbrack S_\beta \,S_1(\nu,Q^2)
+{1\over M^2}\left(\nu\; S-{q\cdot
S}\; P\right)_\beta\,S_2(\nu,Q^2)\right\rbrack\;\; ;\\
&&\qquad\qquad\qquad 
G_i(\nu,Q^2)={1\over 2\pi}{\rm Im}[S_i(\nu,Q^2)]\;\; .
\end{eqnarray}
Owing to the time-ordered product in Eq.~(\ref{Tdef}),
the structure functions $S_i(\nu,Q^2)$ can
in principle be calculated directly in Feynman-Dyson type 
perturbation theory.

In the Bjorken limit, defined by
taking $Q^2,\nu\rightarrow \infty$ while the ratio
$x_B\equiv Q^2/2\nu$ remains fixed, the physics
involved in the structure functions becomes transparent
because the asymptotic freedom of QCD \cite{wilczek}
simplifies the scattering mechanism immensely.
Well-defined relations emerge between the properties of the 
fundamental degrees
of freedom---quarks and gluons---and the experimental data
on the structure functions near this limit.  
The physical significance of the structure functions 
is made more apparant by introducing the scaling functions
\begin{eqnarray}
g_1(x_B,Q^2)&\equiv&{\nu\over M^2}G_1(\nu,Q^2)\;\; ,\\
g_2(x_B,Q^2)&\equiv&\left({\nu\over M^2}
\right)^2G_2(\nu,Q^2)\;\; ,
\label{scaling}
\end{eqnarray}
which remain finite in the Bjorken limit.
It is well-known that $g_1(x_B, Q^2)$ 
contains information about the polarized quark 
and gluon distributions.  From Eqs.~(\ref{decomp}, \ref{scaling}),
one finds that the 
structure function $G_2(\nu,Q^2)$ decouples from the 
scattering cross-section in the Bjorken
limit.  Hence its contribution is formally suppressed by
a power of $1/Q$ in comparison with $G_1(\nu,Q^2)$.  
At tree-level in QCD, $g_2(x_B,Q^2)$ is not related to
a simple parton distribution.  Rather, it contains 
information about quark-gluon correlations within the 
nucleon.  The goal of this paper is to study the effect of
QCD corrections on the relation between $g_2(x_B,Q^2)$ and
these interesting correlations.

At this point, it is convenient to choose the frame
used most frequently in high-energy scattering.
Taking the 3-momenta of $P^\mu$ and $q^\mu$ parallel and
along the $z$-direction,
we write
\begin{eqnarray}
\label{Pexp}
P^\mu&=&p^\mu+{M^2\over2}n^\mu\;\; ,\\
q^\mu&=&-\zeta p^\mu+{Q^2\over 2\zeta} n^\mu\;\; ,
\label{10}
\end{eqnarray}
where the basis vectors $p^\mu$ and $n^\mu$ are
defined by
\begin{eqnarray}
p^\mu&=&(1,0,0,1)\Lambda\;\; ,\\
n^\mu&=&(1,0,0,-1)/2\Lambda\;\; .
\end{eqnarray}
$\zeta$ is given by
\begin{equation}
\zeta={\nu\over M^2}\left\lbrack \sqrt{1+{Q^2M^2\over \nu^2}}
-1\right\rbrack\;\; ,
\label{11}
\end{equation}
and $\Lambda$ is an arbitrary 
scale reflecting our remaining freedom to boost along
the $z$-axis.  Note that this 
basis satisfies $p^2=n^2=0$
and $p\cdot n=1$. 
In the limit $\Lambda\rightarrow\infty$, one
obtains the so-called infinite-momentum frame, in which
Feynman's parton model was originally formulated \cite{feyn}.
In the context of QCD, this limit can serve as the basis for
light-cone power counting, which we will discuss in
detail in the next section.

Along with two independent transverse vectors, $p^\mu$ and $n^\mu$
form a complete basis for Lorentz vectors.
Hence {\it any} vector, $k^\mu$, can be expressed in terms of them :
\begin{equation}
k^\mu=(n\cdot k)p^\mu+(p\cdot k)n^\mu+k_\perp^\mu\;\; .
\end{equation}
In particular, the spin 
polarization, $S^\mu$, has the form
\begin{equation}
S^\mu={h\over M}\left(p^\mu-{M^2\over 2}n^\mu\right)
+\lambda S_\perp^\mu\;\; .
\label{Sexp}
\end{equation}
Here, 
$S_\perp^\mu$ is an arbitrary vector in the 
1- and 2-directions, normalized as 
$S_\perp^2=-1$.
$h$ and $\lambda$ represent the degree of 
longitudinal and transverse polarization, respectively,
and are constrained by $h^2+\lambda^2=1$.
In the limit $\Lambda\rightarrow\infty$, the 
longitudinal polarization
is leading since $p^\mu\sim\Lambda$, whereas 
its transverse components are subleading since
$S_\perp\sim 1$.
Taking $\lambda^2=1$, corresponding to complete
transverse polarization, the 
hadron tensor simplifies to
\begin{equation}
W^{[\mu\nu]}_\perp=-i\epsilon^{\mu\nu\alpha\beta}
q_\alpha \;\lambda\,S^\perp_\beta
\,{M\over\nu}\,\left\lbrack g_1(x_B,Q^2)+
g_2(x_B,Q^2)\right
\rbrack\;\; .
\end{equation}
From the above, it is easy to see that one of the $\mu\nu$ 
indices must be transverse while the other is longitudinal.
More interestingly, it is 
$g_T(x_B,Q^2)\equiv g_1(x_B,Q^2)+g_2(x_B,Q^2)$ which
naturally appears in DIS with a transversely 
polarized nucleon.  For this reason,
we will concern ourselves with this scaling function
and the associated Compton amplitude $S_T(\nu, Q^2) \equiv 
(\nu/M^2)[S_1(\nu,Q^2) + (\nu/M^2)S_2(\nu,Q^2)]$, rather than 
$g_2(x_B,Q^2)$ and $S_2(\nu, Q^2)$. 
 
\section{The Formalism}

In this section, we present a general 
procedure to obtain perturbative QCD predictions for
the structure functions in deep inelastic 
scattering, valid to all orders in $\Lambda_{\rm QCD}/Q$ and 
$\alpha_s(Q^2)$. 
For simplicity, we concern ourselves mainly with the forward
Compton amplitudes. This procedure involves several steps. 
First, one needs to express the amplitudes in terms of 
general parton correlations functions and perturbative scattering
amplitudes. Then a collinear expansion is used to 
simplify the classification of the contributions 
in terms of powers of $\Lambda_{\rm QCD}/Q$. 
In the process, the parton correlation
functions become light-cone correlations which have 
definite power counting properties. 
At a particular order in $1/Q$, certain classes of 
processes differing in the number of
longitudinal gluons can be resummed to form gauge links 
extrapolating between spacially separated physical fields
in our correlations.  This process of resummation is 
greatly simplified by the choice of light-cone gauge in our 
calculations.  Once this choice is made, it becomes 
clear that only a finite number of gauge-invariant
correlations can appear at each given order in $1/Q$.
The coefficients of these correlations 
are themselves power series in the strong coupling, $\alpha_s(Q^2)$.
As with most calculations in quantum field theory, higher
order terms in these expansions contain infinities 
which must be understood and properly handled if
one is to obtain a sensible result.
After all of these steps are understood, one arrives at
a simple set of rules which can be used to 
calculate the contributions to inclusive DIS at 
any order in $1/Q$ and $\alpha_s(Q^2)$.

\subsection{Collinear Expansion, Light-Cone 
Power Counting and Light-Cone Correlations}

As $Q^2\rightarrow \infty$, part of the deep-inelastic scattering
process must be calculable in perturbation theory because of
asymptotic freedom. On the other hand, the nucleon structure
itself is surely non-perturbative. Therefore, the Compton amplitude 
$T^{\mu\nu}$ contains physics both at hard scales, ${\cal O}(Q)$, 
and soft scales, ${\cal O}(\Lambda_{\rm QCD})$. 
In the following discussion, we 
would like to separate the physics at these two scales. Although
most of the discussion in this subsection may be found in the
literature, the way it is presented here is new and general. 

\begin{figure}[t]
\begin{center}
\epsfig{file=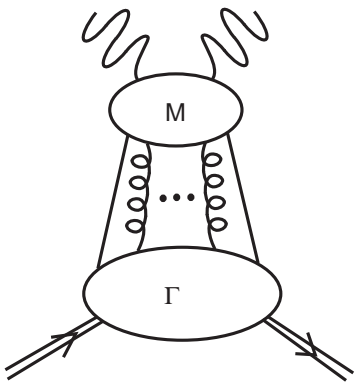,height=5cm}
\end{center}
\caption{Schematic diagram for Compton scattering in terms 
of quark and gluon scattering amplitudes $M$ and their 
correlation functions $\Gamma$ in the nucleon.}
\end{figure}   

As a first step of factorization, we consider the most 
general Feynman 
diagrams consisting of a hard (perturbative) part 
representing scattering 
of a set of quarks and gluons with the photon, and 
a soft (non-perturbative) part representing the 
Green's functions of these quarks and gluons in the
presence of the nucleon, as shown in Fig. 1. 
Denote a collection of incoming
(with respect to the hard scattering) quark momenta by $\{p_i\}$,
outgoing quark momenta by $\{p_j'\}$ and gluon momenta
by $\{k_\ell\}$. The quark and gluon scattering amplitude 
can be expressed as
\begin{eqnarray}
  && M^{\mu\nu} (q, \{p_i\}, \{p_j'\}, \{k_\ell\}) 
  \nonumber\\&&\qquad= i \int e^{iq\cdot\xi} d^4\xi 
   \left\langle 0 \left|{\cal T}\left\lbrace J^\mu(\xi)J^\nu(0)
    \cdots \psi(p_i)\cdots \bar \psi(p_j')\cdots A(k_\ell)\cdots
\right\rbrace\right|
  0\right\rangle_{\rm amp} \ , 
\end{eqnarray}
where the subscript `amp' indicates that 
the external quark and gluon legs are truncated and the 
Lorentz and Dirac indices on gluon and quark fields are
left open. 
The quark and gluon Green's function in the presence
of the nucleon is
\begin{equation}
   \Gamma(\{p_i\}, \{p_j'\}, \{k_\ell\})
  = \left\langle PS\left|{\cal T}\left\lbrace \cdots \bar \psi(p_i) \cdots
  \psi(p_j')\cdots A(k_\ell)\cdots\right\rbrace \right|P\right\rangle \ . 
\end{equation}
Therefore, the full Compton amplitudes can be written as
\begin{equation}
   T^{\mu\nu}(q, p) = 
  \sum_{\rm all ~diagrams}
    \int \prod_{ijl} d^4p_i d^4p_{j'}d^4k_\ell
    M^{\mu\nu}\left(q, \{p_i\}, \{p_j'\},\{k_\ell\}\right)
    \otimes \Gamma\left(\{p_i\}, \{p_j'\},\{k_\ell\}\right) \ , 
\label{couple}
\end{equation}
where the coupling of Lorentz and color indices between
$M^{\mu\nu}$ and $\Gamma$ is implied.
For simplicity, we work with bare (unrenormalized) fields
and couplings until the end of the calculation.

The contributions to $T^{\mu\nu}$ can be classified 
in terms of powers of $1/Q$ (loosely speaking, a twist-expansion). 
Unfortunately, however, 
$1/Q$ power counting is not simply dimensional counting. 
Conceptually, it is easy to understand because
the transverse momentum of a parton, for instance, 
is less important than its longitudinal momentum 
in a hard scattering. We must take these relevant
facts into account systematically through a new 
power counting scheme.
To facilitate this counting, we make a collinear expansion 
of the ``hard 
amplitude"
$M^{\mu\nu}(q, \{p_i\}, \{p_j'\},\{k_\ell\})$ by writing
\begin{eqnarray}
   p_i^\mu&=&x_i p^\mu  + \Delta p_i^\mu\;\; ,\nonumber\\
p_j'\,^\mu &=& x_j' p^\mu  + \Delta p_j'\,^\mu\;\; ,\\
k_\ell^\mu &=& y_\ell p^\mu  + \Delta k_\ell^\mu\;\; ,\nonumber
\end{eqnarray}
where $x_i= p_i \cdot n$,
and similarly for $p_j'$ and $k_\ell$. Expanding
all quark and gluon momenta about the 
collinear point $\Delta p_i^\mu=\Delta p_j'\,^\mu=\Delta k_\ell^\mu=0$,
we obtain 
\begin{eqnarray}
   && M^{\mu\nu}(q, \{p_i\}, \{p_j'\},\{k_\ell\})
  = \sum_{\{m_i\},\{n_j'\},\{t_\ell\}}
    {1\over \cdots m_i!\cdots n_{j}'!\cdots t_\ell!\cdots}
    \nonumber \\
   && \qquad[\cdots (\Delta p_i)^{m_i}\cdots (\Delta p_j')^{n_j'}
    \cdots (\Delta k_\ell)^{t_\ell}\cdots]
    [\cdots \partial \cdots] M^{\mu\nu}(q, 
    \{x_ip\}, \{x_j'p \}, \{y_\ell p\})\;\; ,
\label{expand}
\end{eqnarray}
where we have omitted all the Lorentz indices which couple
the partial derivatives with $\Delta p_i$, $\Delta p_i'$, 
and $\Delta k_\ell$. 
The leading term in the expansion can be interpreted
as Compton scattering off of several 
collinear onshell partons with Feynman 
momentum fractions $x_i=n\cdot p_i$, etc.  Higher order
terms take offshellness and non-collinear effects
into account perturbatively.
  
Rewriting all the quark and gluon momentum 
integrals in Eq.~(\ref{couple})
in terms of these variables, we can 
integrate out $\Delta p_i$, etc. Then the $\Delta$-factors 
in Eq.~(\ref{expand})
become spatial derivatives on the quark and gluon
fields in $\Gamma(\{p_i\},\{p_j'\},\{k_\ell\})$,  
which leads to 
\begin{eqnarray}
   T^{\mu\nu}(q, p) &= &
   \sum_{\rm all ~diag.} \sum_{\{m_i\},\{n_j'\},\{t_\ell\}}
    {1\over \cdots m_i!\cdots n_{j'}!\cdots t_\ell!\cdots}
    \nonumber \\
 && \prod_{ijk} \int dx_i dx_j'dy_\ell
    (\cdots \partial \cdots) M^{\mu\nu}
    (q, p, \{x_i\}, \{x_j'\}, \{y_\ell\})
     \Gamma_{mnt}(\{x_i\},\{x_j'\}, \{y_\ell\})\;\; .
\label{fac2} 
\end{eqnarray}
The new correlations $\Gamma_{mnt}$ have
field separations restricted to the light-cone $n^\mu$-direction
in the coordinate space, and hence are called
light-cone correlations:  
\begin{eqnarray}
  && \Gamma_{mnt}(\{x_i\},\{x_j'\}, \{y_\ell\}) 
  = \prod_{ij\ell}
  \int {d\lambda_i\over 2\pi}
{d\mu_j\over 2\pi}
{d\nu_\ell\over 2\pi} e^{i(\lambda_ix_i+\mu_j x_j'+y_\ell \nu_\ell)}
\nonumber \\
&&
\qquad\qquad\times \langle PS|\cdots\partial^{m_i}\bar\psi(\lambda_i n)
 \cdots \partial^{n_j}\psi(\mu_j n)\cdots 
\partial^{t_\ell}A(\nu_\ell n)\cdots |PS\rangle\;\; ,
\end{eqnarray}
where again we have omitted the Lorentz and color indices.

Now, we are ready to introduce light-cone power counting. 
The hard part in Eq.~(\ref{fac2}) contains only two momenta, 
$q^\mu$ and $p^\mu$, and the only dimensionful scalar 
in the absence of quark mass effects is $Q^2$. 
[The effects of finite quark mass can be taken into
account systematically, as explained in Subsection 3E.]
On the other hand, the soft part depends on the 
scale $\Lambda_{\rm QCD}$. 
Because there are Lorentz indices that couple 
the soft and hard parts, $1/Q$ power counting is 
not determined entirely by the dimension counting of the 
soft and hard part alone. Instead we need a new type of
power counting: light-cone
power counting. Since the $T^{\mu\nu}$ has a fixed 
dimension, we can figure out $1/Q$ counting by studying
behavior of either the hard or the soft part. 
Here we choose to examine the soft part.

The light-cone correlation functions in the collinear expansion
involve QCD fields and their derivatives. Lorentz covariance
requires that these objects be expressed in terms
of $p^\mu$, $n^\mu$, $g^{\mu\nu}_\perp \equiv g^{\mu\nu}
-p^\mu n^\nu - p^\nu n^\mu$, 
$\epsilon^{\mu\nu}_\perp \equiv \epsilon^{\mu\nu\alpha\beta}
p_\alpha n_\beta$, and $S_\perp^\mu$.  
Different $1/Q$ powers result
when these vectors are contracted with the hard partonic
scattering amplitudes. Note that the contractions
of Lorentz structures never 
produce any soft dimension because of the choice
of the collinear coordinates. The balancing soft
dimension in each term comes from the scalar 
coefficients of the expansion, 
taking into account $[p]=1$, $[n]=-1$ 
and $[g]=[\epsilon]=[S_\perp]=0$. Because the dimension
of the Compton amplitude is fixed, the soft dimensions 
of the scalar coefficients determine the associated hard 
dimensions. Obviously, the terms with the
lowest soft dimension dominate the contribution. 

For example, if the soft and hard parts are connected
by two-quark lines, we have the following light-cone 
correlation function :
\begin{equation}
    \int {d\lambda\over 2\pi}\,e^{i\lambda x}
        \langle PS|\bar \psi_\alpha(0)
    \psi_\beta(\lambda n)|PS\rangle
  = 2 \left[ \gamma^\mu_{\beta\alpha} p_\mu 
  q(x) + \gamma^\mu_{\beta\alpha} n_\mu f_4(x)
  + \cdots\right]
\end{equation}
By dimensional counting, $[q(x)]=0$ and $[f_4(x)]=2$.
Therefore, the $q(x)$ contribution to the hard scattering
is leading and the $f_4(x)$ contribution is suppressed
by a factor of $1/Q^2$ relative to $q(x)$. 

The relation between counting the soft dimension and
the use of the infinite-momentum frame is now easy to see. Since
$p^\mu\sim \Lambda$, the contribution of the associated term
is leading in the limit of $\Lambda\rightarrow \infty$. 
On the other hand, $n^\mu\sim \Lambda^{-1}$, so contributions
associated with 
this vector are doubly suppressed. The Lorentz 
structures spanning transverse dimensions are ${\cal O}(\Lambda^0)$
as $\Lambda\rightarrow \infty$. Hence, their contributions 
are suppresed
relative to the $p^\mu$ terms, but enhanced related to 
the $n^\mu$ type.

This fact allows us to separately identify a certain power 
suppression for each field and derivative in the correlation 
function. For gluon fields, one has
\begin{equation}
A^+\sim (1/Q)^0\;\; ,\qquad
A_\perp \sim (1/Q)^1\;\; ,\qquad
A^- \sim(1/Q)^2\;\; \ .  
\end{equation}
As required by the form of the covariant derivative,
partial derivatives obey the same counting rules
as gauge fields.  Now, we see why our power counting arguments
cannot be made until after the collinear expansion : 
subleading momentum components alter the form of the 
correlation functions, leading to an extra suppression.  
Until this behavior is made explicit, our correlation 
functions do not have a definite power counting.
An analysis of the quark bilinears, taking into account the fact
that fermion fields have mass dimension 3/2, leads
to
\begin{eqnarray}
\overline\psi\gamma^+(1, \gamma_5)\psi \sim (1/Q)^2\;\; ,\qquad
\overline\psi\gamma^\perp(1, \gamma_5)\psi \sim(1/Q)^3\;\; ,\qquad
\overline\psi\gamma^-(1, \gamma_5)\psi\sim(1/Q)^4\nonumber\;\; ,\\
\overline\psi(1,\gamma_5)\psi \sim (1/Q)^3\;\; ,\qquad\qquad
\overline\psi\sigma^{+-}\psi \sim (1/Q)^3\;\; ,\qquad\qquad
\overline\psi\sigma^{\perp\perp '}\psi \sim (1/Q)^3\;\; ,
\qquad
\end{eqnarray}
where $\gamma^\pm = (\gamma^0\pm \gamma^3)/\sqrt{2}$ and $(1,\gamma_5)$ means either 1 or $\gamma_5$. 
These relations can be simplified by breaking the 
quark field into two parts :
\begin{equation}
\psi =\psi_++\psi_-\;\; ,\qquad
\psi_\pm = {1\over2}\gamma^\mp\gamma^\pm\psi\;\; ,
\end{equation}
and writing simply
\begin{equation}
\psi_+\sim (1/Q)^1\;\; ,\qquad
\psi_-\sim (1/Q)^2\;\; .
\end{equation}
The suppression associated with an arbitrary correlation
function is simply the sum of the suppressions
associated with its constituent fields and derivatives minus 
two for the external nucleon states.  

A general light-cone correlation $\Gamma$ is 
the nucleon matrix element
of an operator with $n_{\phi}$ $\phi$ 
fields, for $\phi=A^+, A^\perp,
A^-, \psi_+, \psi_-$, and $n_{\theta}$ partial derivatives,
for $\theta=\partial_\perp, \partial^-$. According to
the above discussion, its soft
dimension is given by
\begin{equation}
n_{\partial_\perp}+n_{A_\perp}+n_{\psi_+}
+2(n_{\partial^-}+n_{A^-}+n_{\psi_-})+0\cdot n_{A^+}-2\;\; .
\end{equation}
Armed with this formula, we can immediately 
determine the order at which a particular 
correlation function enters our amplitudes.
For instance, the leading contribution
to DIS comes from correlations with either two 
$\psi_+$ or two $A_\perp$ correlations with
arbitrary number of $A_+$. 

For polarized DIS with transverse polarization, the leading
contribution is of order $1/Q$. The possible field combinations
at this order are $\bar \psi_\mp\psi_\pm$, 
$A_-A_\perp$, $A_\perp A_\perp A_\perp$, $\psi_+ A_\perp
\psi_+$, $\psi_+\partial_\perp \psi_+$ and $A_\perp \partial_\perp
A_\perp$, each appearing with an 
arbitrary number of $A^+$'s. For each of these
correlations, we need to consider all possible Feynman 
diagrams with corresponding external legs. The momenta
entering these legs are all collinear.

\subsection{Longitudinal Gluons, Gauge Invariance and the Light-Cone 
Gauge}

As we have seen in the last subsection, the 
longitudinal gauge potential, $A^+$, in 
light-cone correlations does not lead to any suppression 
in its contribution to scattering amplitudes. 
At any given order in $1/Q$, an infinite set of 
correlations with increasing number of $A^+$'s contribute. 
A related problem is that of gauge-invariance.
The correlations that naturally appear in the collinear
expansion are not manifestly gauge invariant.
In particular, the spatially separated fields 
do not form gauge-invariant operators without
appropriate gauge links.
The appearance of gauge potentials also poses a gauge-invariance
problem since these objects do no transform covariantly under 
the action of the gauge group.
As we have mentioned before, 
the parton scattering amplitudes $M^{\mu\nu}$ 
are calculated with external collinear and 
on-shell parton states after collinear expansion. 
As such, they are separately gauge invariant. Since
the Compton amplitudes are gauge invariant order by order
in $1/Q$, all correlations at a given order 
must combine in such way to yield gauge invariant
parton distributions. Gluon potentials and 
partial derivatives do not transform covariantly 
under the gauge group, so they must combine to 
form either covariant derivatives or field 
strength tensors.

In this subsection, we argue that 
the infinite number of correlations 
must be combined into a finite number of gauge-invariant parton 
distributions. In particular,
the correlations with increasing numbers of $A^+$'s
form gauge-invariant structures with gauge links 
connecting fields at separate spacetime points.
This conclusion is, of course, 
well-known in the case of leading twist \cite{cfp}. 
We assert that it is in fact true order
 by order in $1/Q$.

\begin{figure}[t]
\begin{center}
\epsfig{file=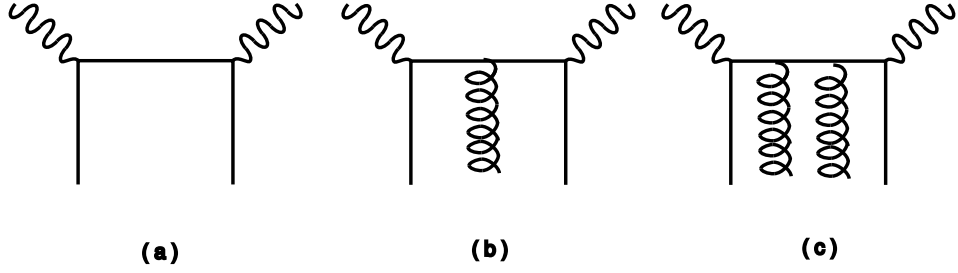,height=3.5cm}
\end{center}
\caption{A set of tree-level twist-two contributions to  
DIS, differing in the number
of longitudinal gluons participating in hard scattering.}
\end{figure}   

Let us recall the argument of gauge-link formation 
in the leading twist case.  Consider a class of diagrams 
with a number of $A^+$ gluons connecting the soft and hard parts.
As we stated above, the hard part is already gauge invariant and can
be calculated in any gauge one likes. 
Using Ward identities derived from the gauge symmetry,  
one factorizes all the longitudinal gluon 
lines connected to the hard
scattering blob onto eikonal lines. 
These eikonal lines represent the gauge links. 
To prove this fact, one 
refers to the proof of factorization by Collins, 
Soper and Sterman in Ref.~\cite{mueller}. At leading 
order in $1/Q$ and $\alpha_s(Q^2)$, 
this process is especially easy to see.  
Collecting the contributions from the 
diagrams in Fig. 2, one obtains
\begin{eqnarray}
T^{\mu\nu}&=&ie_q^2\int\,{dxd\lambda\over2\pi}\,e^{i\lambda x}
\left\langle PS\left|\bar\psi_+(0)\gamma^\mu\right.\right.\nonumber\\
&&\qquad\times\left\lbrack
{i\over x\not\!p+\not\!q}+\int\,{dy_1 d\mu_1\over2\pi}
\,e^{i\mu_1 y_1}\,{i\over (x+y_1)\not\!p+\not\!q}
\left(-igA^+(\mu_1 n)\gamma^-\right)
\,{i\over x\not\!p+\not\!q}\right.\nonumber\\
&&\qquad\left.+\int\,{dy_1 d\mu_1\over2\pi}\,{dy_2 d\mu_2\over2\pi}
\,e^{i\mu_1 y_1+i\mu_2 y_2}\,{i\over (x+y_1+y_2)\not\!p+\not\!q}
\left(-igA^+(\mu_2 n)\gamma^-\right)\,\right.\nonumber\\
&&\qquad\qquad\times\left.
{i\over (x+y_1)\not\!p+\not\!q}
\left(-igA^+(\mu_1 n)\gamma^-\right)
\,{i\over x\not\!p+\not\!q}+\cdots\right\rbrack
\left.\left.\gamma^\nu
\psi_+(\lambda n)\right|PS\right\rangle\;\; \\
&&\qquad\qquad\qquad\qquad
\qquad\qquad+(\mu\leftrightarrow\nu,q\rightarrow-q)\;\; .\nonumber
\end{eqnarray}
For illustrative purposes, we take 
$\mu\nu$ in the transverse dimensions
and simplify the Dirac algebra.  Giving 
$x_B = Q^2/2p\cdot q$ a small negative imaginary 
part and 
performing the integrals over the $y_i$, we obtain
\begin{eqnarray}
T^{\mu\nu}&=&-{e_q^2\over2}\int{dxd\lambda\over2\pi}
\;{1\over x-x_B}\;e^{i\lambda x}\left\langle PS\left|
\bar\psi(0)\gamma^\mu\not\!n\right.\right.\nonumber\\
&&\times\left\lbrack
1+\int_\lambda^0 d\mu_1\,(-igA^+(\mu_1 n))
+\int_\lambda^0 d\mu_1\,(-igA^+(\mu_1 n))\int_\lambda^{\mu_1}
d\mu_2\,(-igA^+(\mu_2 n))+\cdots\right\rbrack\nonumber\\
&&\qquad\qquad\qquad\times\left.\left. \gamma^\nu\psi(\lambda n)
\right|PS\right\rangle\;\; +\;\;(\mu\leftrightarrow\nu,
q\rightarrow-q)\;\; .
\end{eqnarray}
The quantity in brackets is precisely the 
gauge link required to make this distribution
gauge-invariant.  

At higher orders in the $1/Q$ expansion, 
we have more fields and this process of
resummation becomes more complicated.
However, at any given order in $1/Q$, 
there is only a finite number of $A_\perp$ and $A^-$
according to power counting. These potentials 
require only a finite number of $A^+$'s and partial 
derivatives to form gauge covariant quantities.
Therefore the infinite number of remaining $A^+$'s 
must form gauge-covariant objects by themselves. One can 
easily see by inspection that this can only 
be achieved if they form gauge links.
Once done, we are left with a finite number
of correlations with gauge links.

After an infinite number of longitudinal gluons
have been resummed to form the gauge links, 
the remainder must combine with the partial derivatives and 
the $A_\perp$ and $A^-$ fields to form
fully gauge-invariant distributions.  
This means that they must organize into covariant derivatives
and field strength tensors.  Although difficult to prove, 
this fact is a direct consequence of gauge invariance.
It can be used as a check of any explicit calculation.

In a practical calculation, the presence of the
longitudinal gluons represents an unncessary
complication. The standard approach to avoiding
this is to make the choice of light-cone gauge
$A^+=0$. Although a gauge choice here is
not required since we are not making perturbative
calculations of the correlations, it represents a
tremendous simplification. 
In this gauge, no correlations involving $A^+$
are ever needed and the gauge links simply
reduce to unity. The only complication comes at the 
end of the calculation, when we would like to organize
the correlations in the
axial gauge into manifestly
gauge invariant ones.

As a last point in this subsection, we mention that 
not all the correlations that one can write
down according to the light-cone power counting above
are independent. The so-called `bad' field components,
$\psi_-$ and $A^-$, are related to 
the `good' components, $\psi_+$ and $A_\perp$,
through the QCD equations of motion:
\begin{equation}
\not\!n\; \not\!\!\!\!iD\psi=\not\!n\!\not\!p\, iD^+\psi
+\not\!n \;\not\!\!\!\!iD_\perp\psi
=2iD^+\psi_-+\not\!n\;\not\!\!\!\!iD_\perp\psi_+=0\;\; ,
\end{equation} 
where we have introduced the covariant derivative
\begin{equation}
D_\alpha(\mu n)\equiv \partial_\alpha+igA_\alpha(\mu n)\;\; .
\end{equation}
In the light-cone gauge, we can solve this equation for the
`bad' quark field component in terms of the 
`good' one:
\begin{equation}
\label{badq}
\psi_-(\lambda n)={i\over 2}\int^{\lambda}\,d\zeta
\not\!n\;\not\!\!\!\!iD_\perp(\zeta n)\psi_+(\zeta n)\;\; .
\end{equation}
Similarly, one 
has
\begin{equation}
\label{badg}
A_a^-(\lambda n)=-\int^\lambda\,d\zeta\,\int^\zeta\,
       d\zeta'\left\lbrack
D^{ab}_\perp(\zeta' n)\partial^+ A_b^\perp(\zeta' n)+
g\overline\psi_+(\zeta' n)\not\!n\,t^a\psi_+(\zeta' n)\right\rbrack 
\ . 
\end{equation}
from the gluon equation of motion. 
The above equations are consistent with the light-cone power 
counting of the previous subsection. 
In both cases, we see that the `bad' components, $\psi_-$ and $A^-$,
can be eliminated in favor of either two `good' fields, $\psi_+$ or
$A_\perp$, or one `good' field and a transverse momentum insertion.
This follows from the fact that in light-cone quantization,
`bad' field components do not propagate freely and hence cannot 
be directly interpreted in Feynman-Dyson type 
perturbation theory. As a consequence, we can choose
independent correlations without inclusion of any 
``bad'' fields.

\subsection{Example: $S_T$ in the Leading Order}

As an example of formalism presented above, we 
consider the leading contribution to $S_T(\nu,Q^2)$.  
This example is useful not only as a simple 
illustration of the procedure, but 
also to familiarize the reader with some of the  
distributions that appear in DIS with transverse
polarization. To simplify the discussion, we contract
one of the photon polarizations with $n^\mu$.  This 
leads to
\begin{equation}
T^{[\mu+]}=i\epsilon^{-+\beta\mu}\,
{MS_\beta\over\nu}\,x_B\,S_T(\nu,Q^2)\;\; ,
\end{equation}
where $\mu$ is transverse.

In the light-cone gauge,
there are two independent correlations at leading
order in $\alpha_s(Q^2)$. The first involves the good
components of two quark fields and a transverse 
partial derivative:
\begin{eqnarray}
    \Gamma_{qqB}^\alpha(x,y) &=& \int {d\lambda \over 2\pi}
      {d\mu\over 2 \pi} e^{i\lambda x}
      e^{i\mu(y-x)}\langle PS|\bar \psi_+(0)
     i\partial^\alpha_\perp \psi_+(\lambda n) 
       |PS\rangle \nonumber \\
     & = & \delta(x-y) \int {d\lambda\over
     2\pi}e^{i\lambda x}\langle PS|\bar \psi_+(0)
    i\partial^\alpha_\perp \psi_+(\lambda n) |PS\rangle \ , 
\end{eqnarray}
where the Dirac and color indices on quark fields are open
and $\alpha$ is transverse. The second correlation contains 
two quark and one gluon field : 
\begin{equation}
    \Gamma_{qgqB}^\alpha(x,y) = \int {d\lambda \over 2\pi}
      {d\mu\over 2 \pi} e^{i\lambda x}
      e^{i\mu(y-x)}\langle PS|\bar \psi_+(0)
     (-g_B)A^\alpha_\perp(\mu n) \psi_+(\lambda n) 
       |PS\rangle \ . 
\end{equation}
The leading-order result for $S_T(\nu, Q^2)$ 
must be expressible in terms of these correlations alone.

\begin{figure}[t]
\begin{center}
\epsfig{file=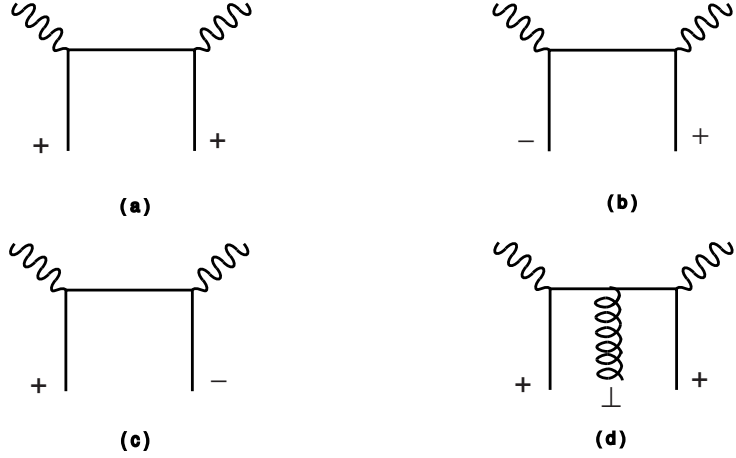,height=6.0cm}
\end{center}
\caption{Tree-level Feynman diagrams for $S_T$. The pluses (+)
and minuses (-) refer to the light-cone projections
of the Dirac fields.}
\end{figure}   

The tree level Feynman diagrams for $M^{\mu+}$ are shown 
in Fig. 3. The total contributions to $T^{\mu+}$
can be expressed as 
\begin{eqnarray}
T^{[\mu+]} = &&{e_q^2\over2\nu}\int\,dx\,dy
\int\,{d\lambda\over2\pi}\,{d\mu\over2\pi}
\,e^{i\lambda x}\,e^{i\mu(y-x)}\nonumber\\
\label{special}
&&\times\,\left\langle PS\left|\overline\psi_+(0)\left\lbrack
{1\over y-x_B}\,\gamma^\mu\not\!n\; \not\!\!\!\!iD_\perp(\mu n)
-{1\over x+x_B}\;\not\!\!\!\!iD_\perp(\mu n)\not\!n\gamma^\mu
\right\rbrack\,\psi_+(\lambda n)\right|PS\right\rangle\\
&&-{e_q^2\over2\nu}\int\,dx\int{d\lambda\over2\pi}\,e^{i\lambda x}
\left\langle PS\left|
\overline\psi_-(0)\not\!p\not\!n\gamma^\mu\psi_+(\lambda n)
\right|PS\right\rangle\\
&&+{e_q^2\over2\nu}\int\,dx\int{d\lambda\over2\pi}\,e^{i\lambda x}
\left\langle PS\left|
\overline\psi_+(0)\gamma^\mu\not\!n\not\!p\,\psi_-(\lambda n)
\right|PS\right\rangle\;\; .
\end{eqnarray}
The special form of the kernels in contribution~(\ref{special})
allows us to ``simplify'' this contribution using the 
equation of motion.  It then combines with the others 
to give the full result
\begin{eqnarray}
T^{[\mu+]}&=&i\epsilon^{\mu+-\alpha}(-x_B)\,{e_q^2\over2\nu}\,\int
\,dx\,\left({1\over x-x_B}-{1\over x+x_B}\right)\nonumber\\
&&\qquad\times\int\,{d\lambda\over2\pi}e^{i\lambda x}\,
\left\langle PS\left|\overline
\psi(0)\gamma_\alpha\gamma_5\psi(\lambda n)
\right|PS\right\rangle\;\; .
\end{eqnarray}
As a consequence of parity and time-reversal invariance, 
one can isolate a scalar distribution,
\begin{equation}
\int{d\lambda\over4\pi}\,e^{i\lambda x}\,
\left\langle PS\left|\overline\psi(0)\gamma_\alpha
\gamma_5\psi(\lambda
n)\right|
PS\right\rangle=(MS)_\alpha\,q_{TB}(x)\;\; ,
\end{equation}
and write 
\begin{equation}
\label{mislead}
S^{(0)}_T(\nu,Q^2)=-\sum_q
e_q^2\int\,dx\left({1\over x-x_B}-{1\over x+x_B}\right)
q_{TB}(x)\;\; 
\end{equation}
at leading order in the strong coupling.  Recall that this
distribution, as it appears here, is not renormalized.  However, 
since the difference appears only at higher orders
in $\alpha_s(Q^2)$, we can replace it with a renormalized 
distribution $q_T(x,Q^2)$. [Note that $q_T(x,Q^2)$ can be made
manifestly gauge-invariant 
by re-introducing the $A^+$ fields in the form
of a light-cone gauge link connecting 
the quark fields.]

It is now a simple matter to determine the 
scaling function $g_T(x_B,Q^2)$ in terms of $q_T(x,Q^2)$ :
\begin{equation}
g^{(0)}_T(x_B,Q^2)={1\over 2}\sum_q\,e_q^2\,
\left\lbrack q_T(x_B,Q^2)+q_T(-x_B,Q^2)\right\rbrack\;\; .
\end{equation}
This extremely simple result is quite misleading.
It seems to imply that only one function of a single 
variable is required to predict the ${\cal O}(1/Q)$
corrections to DIS.  In analogy with the leading ${\cal O}(1)$
result, one might assume that higher order corrections
will do nothing but modify the form of the coefficient function.
Furthermore, one might expect that the $Q^2$ variation
of this new distribution is fixed from the distribution itself.
Unfortunately, neither of these suspicions is correct.
The fact that our leading result can be expressed so simply in terms
of $q_T(x,Q^2)$ is due exclusively to the form
of the coefficient function in contribution~(\ref{special}).  
There is no reason to believe that this
simple form will persist at higher orders.  

To reveal the full content of $g_T(x_B, Q^2)$ at this order,
we use the equation of motion to eliminate the ``bad''
fields. The result can be expressed in terms of the 
single gauge-invariant correlation
\begin{eqnarray}
      K_{qB}^\alpha(x, y)  &=& 
\Gamma^\alpha_{qqB}(x,y)+\Gamma^\alpha_{qgqB}(x,y)\nonumber\\
&=&\int {d\lambda \over 2\pi}
      {d\mu\over 2 \pi} e^{i\lambda x}
      e^{i\mu(y-x)}\langle PS|\bar \psi(0)
     iD^\alpha_\perp(\mu n) \psi(\lambda n) 
       |PS\rangle \ ,
\end{eqnarray}
where color indices have been contracted but 
the Lorentz indices are open. The only invariant
correlation $K_{qB}(x, y)$ appears in the 
contraction  
\begin{eqnarray}
\int\,{d\lambda\over4\pi}\,{d\mu\over4\pi}\,e^{i\lambda x}
\,e^{i\mu (y-x)}\,\left\langle PS\left|
\overline\psi(0)\;\not\!\!\!\!iD_\perp(\mu n)\not\!n\gamma^\mu
\psi(\lambda n)\right|PS\right\rangle=
\phantom{-}i\epsilon^{-+\alpha\mu}MS_\alpha K_{qB}(x,y)
\;\; ,\\
\int\,{d\lambda\over4\pi}\,{d\mu\over4\pi}\,e^{i\lambda x}
\,e^{i\mu (y-x)}\,\left\langle PS\left|
\overline\psi(0)\gamma^\mu\not\!n\;\not\!\!\!\!iD_\perp(\mu n)
\psi(\lambda n)\right|PS\right\rangle=
-i\epsilon^{-+\alpha\mu}MS_\alpha K_{qB}(y,x)\;\; ,
\end{eqnarray}
of these latter indices.
Since 
\begin{eqnarray}
\int\,{d\lambda\over4\pi}\,{d\mu\over4\pi}\,e^{i\lambda x}
\,e^{i\mu (y-x)}\,\left\langle PS\left|
\overline\psi(0)(-ig)F^{+\alpha}(\mu n)\gamma_\alpha\not\!n\gamma^\mu
\psi(\lambda n)\right|PS\right\rangle\nonumber\\
\qquad\qquad\qquad\qquad=
i\epsilon^{-+\alpha\mu}MS_\alpha (y-x)K_{qB}(x,y)\;\; ,
\label{Fpart}
\end{eqnarray}
every gauge-invariant way to include two `good' quark
fields and either a `good' gluon field or a transverse partial 
derivative can be accomodated by $K_{qB}(x,y)$.  Furthermore,
the first moment of $K_{qB}(x,y)$ with respect to $y$ 
generates the `bad' component of the quark field
through the equation of motion.  Hence all 
of the distributions relevant to ${\cal O}(1/Q)$
DIS are contained in $K_{qB}(x,y)$.  This fact allows us
to assert that {\it all} ${\cal O}(1/Q)$ corrections
to DIS in the nonsinglet sector can be expressed in terms
of $K_{qB}(x,y)$.  In the singlet sector, one must introduce one
more distribution to describe pure gluonic effects.  This is
done in Section 4.  

In terms of $K_{qB}(x,y)$, our leading correction has the form
\begin{equation}
S^{(0)}_T(\nu,Q^2)=-\sum_qe_q^2\int\,dx\,dy\,{2\over x}\left(
{1\over x-x_B}-{1\over x+x_B}\right)\,K_q(x,y,Q^2)\;\; ,
\end{equation}
where we have once again ignored the difference between 
renormalized and unrenormalized distributions.
In the remainder of this paper, we will be concerned 
with determining the ${\cal O}(\alpha_s)$ corrections 
to this result.

\subsection{Factorization of Infrared 
Singularities and Radiative Corrections}

Up to this point, we haven't said much about the
perturbative coefficients $M^{\mu\nu}$ of the correlation
functions other than that they are calculated
in terms of collinear and on-shell external
parton states and that they are gauge invariant.
They are, of course, perturbative expansions
in the strong coupling, $\alpha_s(Q^2)$. Hence, the final 
result for $T^{\mu\nu}$ is a double expansion. 
As $Q$ becomes large, both expansion parameters
vanish. Loosely speaking, the $1/Q$ expansion is associated
with the number of fields in the soft part of the process, 
while the coupling expansion is associated with those in the 
hard part. In this subsection, we would like to
elucidate the nature of the latter expansion.

Because the partons are on-shell, $M^{\mu\nu}_i$ can be 
viewed as a parton scattering S-matrix element. In principle, 
one must multiply by parton wave function renormalization 
factors to get the proper S-matrix element. 
However, working in $d=4-\epsilon$ dimensions, the absence
of a physical scale at the massless poles 
of the quark and gluon propagators reduces these
contributions to unity.
Subleading terms in the expansion of $M^{\mu\nu}_i$
are parton scattering S-matrix elements with insertions
of certain vertices associated with
powers of $k_j^\perp$ and $k^-_j(\equiv p\cdot k_j)$. 
For instance, with one power of quark 
momentum $k_j^\perp$, the subleading term 
is calculated with one insertion of the vector
vertex $i\gamma^\alpha_\perp$ to one of the quark propagators.
Because S-matrix elements and their relatives are 
gauge invariant, one may choose any gauge for the internal
gluon propagators in $M^{\mu\nu}_i$. We use 
Feynman's choice in our calculation.

As we have argued above, $M_i^{\mu\nu}$ is
ultraviolet finite because the on-shell 
wave function renormalization is trivial in dimensional
regularization. Nevertheless, due to the massless
on-shell external states, $M_i^{\mu\nu}$ has
infrared divergences showing up as $1/\epsilon$ poles. 
To understand the origin of the infrared divergences, 
we consider a vertex correction to the tree-level Compton
scattering diagram. Let the gluon momentum be $k$
and one of the quark propagators $k+xp$. 
Then the loop integral is divergent when $k$ is parallel to
$xp$. Using dimensional regularization, this
collinear divergence appears
as $1/\epsilon$ pole. These 
divergences may be factorized in the perturbative sense  
\begin{equation}
      M^{\mu\nu}_i = C^{\mu\nu}_i \otimes P_i \ ,
\label{52} 
\end{equation}
where $C_i^{\mu\nu}$ is the finite coefficient function 
and $P_i$ contains only the $1/\epsilon$ poles.
On the other hand, the infrared-finite quantities $\Gamma_{iB}$ 
contain ultraviolet divergences which also show 
up as $1/\epsilon$ poles. When the infrared poles 
in $P_i$ cancel all the ultraviolet poles 
in $\Gamma_{iB}$, $T^{\mu\nu}$ is said to be 
factorizable.  The product $P_i\Gamma_{iB}$
defines the renormalized parton correlation 
functions $\Gamma_i$. The final factorization formula
for the Compton amplitude is then
\begin{equation}
      T^{\mu\nu} = \sum_i C^{\mu\nu}_i\otimes \Gamma_i\  , 
\end{equation}
where $C^{\mu\nu}_i$ is a well-defined 
perturbation series in $\alpha_s$
and $\Gamma_i$ is a finite nonperturbative distribution.

The Feynman diagrams in the hard part 
are simpler to evaluate after the collinear expansion 
because all the external momenta become collinear
($p$ and $q$).  The only transverse momentum components
come from the loop momenta. Therefore, it is convenient 
to introduce the light-cone coordinates
\begin{eqnarray}
    k^\pm = {1\over \sqrt{2}}(k^0\pm k^3); ~~~
    k_\perp = (k^1, k^2) \ . 
\end{eqnarray}
The momentum-integral now becomes 
$d^4k = dk^+dk^-d^2k_\perp$. The $dk^-$
integral can be done using the residue theorem 
by considering the complex plane of $k^-$.  After this is
done, the transverse $k_\perp$ integrals become straightforward.
The remaining $k^+$ integral is, in general, a nontrivial
multidimensional integral containing both the 
evolution kernels $P_i$ and the finite coefficient functions
$C_i^{\mu\nu}$.  Although this integral can be quite complicated, its
dependence on $\epsilon$ has been made explicit 
by performing the transverse integrals.  This allows us
to separate the process of verifying factorizability from the 
calculation of the coefficient functions.

\subsection{Summary}

We are now in a position to write down a set of rules for
calculating the Compton amplitudes at an arbitrary order, $n$,
in $1/Q$.  First, one must calculate the 
hard scattering Compton amplitude for every combination of 
$g$ `good' field components, where $g$ runs from 2 to $n+2$.
Only diagrams in which $q^\mu$ flows through every
loop are considered in this calculation.  
Initial and final state interactions such as those represented
in Fig. 4 are taken into account via renormalization $Z$-factors 
and the inclusion of the `bad' field components.
The external momenta in these amplitudes are written as
\begin{equation}
\label{momexp}
     k_j^\mu = (k_j\cdot n) p^\mu + \lambda^2(k_j\cdot p) n^\mu 
+ \lambda k_j^\perp \;\; ,
\end{equation}
and the amplitudes are expanded about $\lambda=0$.
The coefficient of 
$\lambda^{n+2-g}$ is convoluted with the 
correlation function, and factors of $k_j^\perp$ and $k_j^-$
are eliminated in favor of partial derivatives acting on fields.
Second, one considers processes which involve one `bad' field 
component and anywhere from one to $n$ `good' field components.
The above process is repeated, but here the relevant coefficient
comes at order 
$\lambda^{n+2-g-2}$.  Next, amplitudes with two `bad' field components
must be considered.  This process continues until we have exhausted
all possible ways to obtain a suppression of 
$(\Lambda_{\rm QCD}/Q)^n$.  In general,
one must consider a collinear expansion to order 
$\lambda^{n+2-g-2b}$ of all amplitudes containing $g$ external
`good' field components and $b$ external `bad' field components
for which $n+2-g-2b$ is positive or zero.   

\begin{figure}[t]
\begin{center}
\epsfig{file=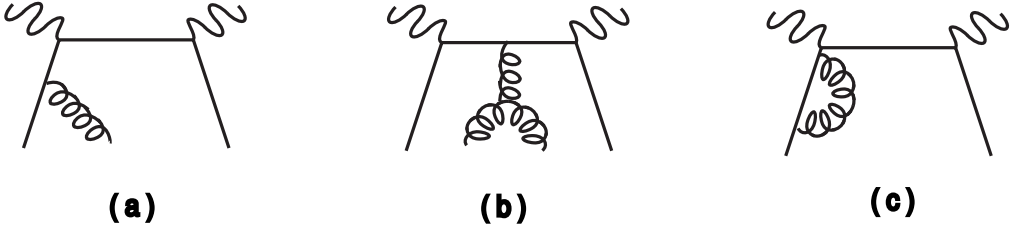,height=3.0cm}
\end{center}
\caption{Hard scattering diagrams which represent processes that
occur either entirely before or after the hard scattering.
$(a)$ and $(b)$ are taken care of by the inclusion of
`bad' field components along with `good' ones, while
the contribution from $(c)$ is grouped into a renormalization
$Z$-factor.}
\end{figure}   

As exemplefied by Eqs.~(\ref{badq},\ref{badg}), we could choose to 
eliminate the `bad' components in the beginning of the calculation.
In this scheme, some diagrams representing processes that 
occur entirely either before or after the hard scattering, such as
Figs. 4$(a)$ and $(b)$ (but not $(c)$), are reinstated.  Rules for the 
onshell couplings and propagators that appear in these
special diagrams are derived directly from the
equations of motion.  This procedure is used in 
Refs. \cite{cfp,ournonsing}.  Throughout the 
rest of this paper, however, we opt for the former method.

The dependence of $q^\mu$ on the nucleon mass
$M$ through $\zeta$ (Eqs.~(\ref{10},\ref{11})) prompts us to 
replace $M\rightarrow \lambda M$ in $\zeta$ before each 
expansion.  This produces additional suppression
of certain correlations, but does not significantly
alter the above analysis because its effects
can always be grouped into the physical vector
$q^\mu$.
If quark mass effects
are desired, one has only to replace 
$\;\not\!\!\!\!iD_\perp$ with $\;\not\!\!\!\!iD_\perp
-m_q$ in Eq.~(\ref{badq}), 
consider $m_q\rightarrow\lambda m_q$ in the amplitudes,
and expand as before.  

We note here that these expansions may be 
asymptotic in nature and
cannot be used to predict experiment to 
arbitrary precision because of the problems presented by infrared 
renormalons \cite{renormalon}. 

\section{One-Loop Corrections to the Transverse Compton amplitude $S_T$}

Armed with the formalism presented in the last section,
we are ready to study the main subject of the paper:
one-loop corrections to the $g_T$ scaling function. 
As we have explained before, it is the Compton
amplitude $S_T$ that is a natural object to
calculate in Feynman-Dyson perturbation theory;
we focus on it exclusively throughout this section.
Part of the results here have already appeared
in our previous publications \cite{ournonsing,oursing}. 
More details are given here for the reader's convenience.

\subsection{One-Loop Corrections in the Non-Singlet Sector}

The non-singlet sector requires at least two quark 
fields, so the correlation functions we need to 
consider are exactly the same as those in 
Section 3C. Before we begin, however, some 
comments are in order.

As with most quantum loop-corrections, our calculation
contains divergences.  As explained above
Eq. (\ref{52}), these divergences 
are of infrared origin after usual renormalization
and can be combined with the bare distributions to 
generate a finite, scale-dependent 
renormalized distribution. Working
in dimensional regularization, one can isolate these 
divergences and obtain a relation between the renormalized 
and bare distributions.  This relation allows us to determine
the $Q^2$ evolution of $K_q(x,y,Q^2)$ and check the 
validity of factorization for our process.  Of course, 
in order to use dimensional regularization to calculate
a spin-dependent quantity, one must specify a 
prescription for $\gamma_5$ and $\epsilon^{\mu\nu\alpha\beta}$.
In this paper, we choose `t Hooft and Veltman's 
prescription \cite{thooft}.

In principle, to obtain the full amplitude one must 
calculate all relevent Feynman diagrams.  
However, loops associated exclusively with external legs
can be grouped into renormalization $Z$-factors
for the external fields.  The absence of a physical scale at the 
massless poles of the physical external propagators 
reduces these factors to unity.
In addition, when both photon
indices are left open, fundamental symmetries allow
one to reduce the number of diagrams that
actually need to be calculated.  The crossing symmetry, 
for example, allows one to cut independent diagrams in
half by asserting symmetry under the replacement
$(\mu\leftrightarrow\nu, q\rightarrow-q)$.  
In our case, the freedom to exchange
$\mu$ and $\nu$ is destroyed by our asymmetric treatment of these
indices.  However, we know that the part of the amplitude 
we search for is antisymmetric in $\mu$ and $\nu$, 
so we can obtain the full result simply by calculating
all diagrams with a certain ordering of photon insertions
and subtracting $(q\rightarrow-q)$ in the end.
Although these diagrams are not all independent 
when photon indices are left open, they represent the 
minimal set of independent contributions to our calculation.

By inspection, one can see that there are four diagrams associated
with $(g,b)=(2,0)$, sixteen associated with $(3,0)$, and 
eight with $(1,1)$.  These diagrams can easily be grouped according 
to color factor.  All of the diagrams associated with $(2,0)$ and $(1,1)$
are proportional to $C_F=(N^2-1)/2N$ in SU$(N)$, while 
those representing $(3,0)$ also include the factors $C_A=N$ and 
$C_F-C_A/2=-1/2N$.  Routing the transverse momentum through the
quark line and performing the collinear
expansion 
\begin{equation}
{i\over\not\!\ell+\not\!k_\perp}={i\over\not\!\ell}+{i\over\not\!\ell}
\;\not\!\!\!ik_\perp\,{i\over\not\!\ell}+\cdots
\end{equation}
before evaluation, we see that the four (2,0) graphs actually 
represent 
twelve contributions, each with an insertion of the 
operator $\;\not\!\!\!ik_\perp \rightarrow 
i\;\not\!\!\!i\partial_\perp$, where the partial derivative
acts on the external quark field.  
Each of these contributions corresponds to a similar 
graph in which the operator insertion is replaced by a gluon line.
The only differences between these contributions are the change
in momentum induced by a gluon absorption and the associated
gauge group generator.  Since the generator contributes
only to the color factor, the $(2,0)$ contribution is simply
the $C_F$-part of the $(3,0)$ contribution evaluated at zero 
momentum transfer, i.e. $x=y$.  This implies that the sum
of $(2,0)$ and $(3,0)$ is obtained by replacing $-gA_\alpha(\mu n)$
with $iD_\alpha(\mu n)$ in the $C_F$-part of $(3,0)$.
Hence, the $C_F$-part of our amplitude is already expressible
in terms of $K_{qB}(x,y)$ alone.  Since there are no 
partial derivative
terms proportional to $C_A$, this part must be gauge-invariant by 
itself.  Looking at Eq.~(\ref{Fpart}), we see that the 
necessary requirement is that its coefficient vanish 
at least linearly as $x\rightarrow y$.
We will see below that this is in fact the case.

\begin{figure}[t]
\begin{center}
\epsfig{file=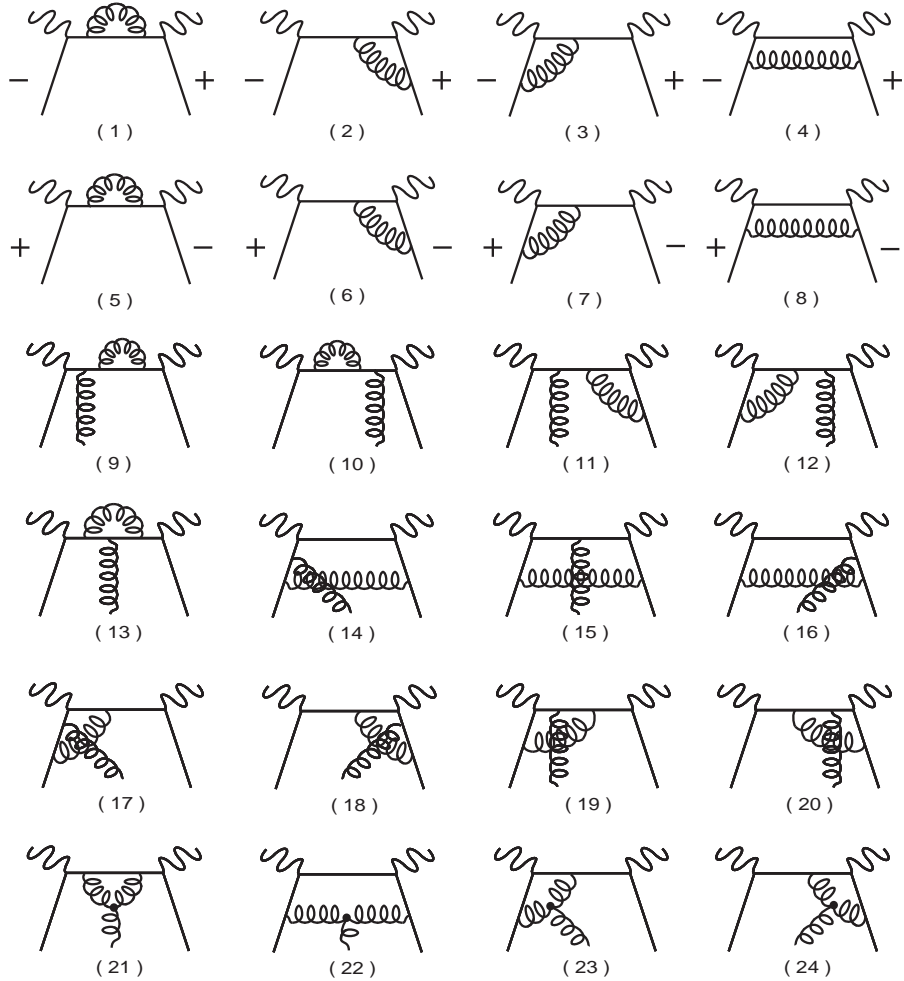,height=13.cm}
\end{center}
\caption{Diagrams contributing to the hard scattering
coefficient $M^{\mu+}$ relevant to $S_T$ at one-loop order
in the non-singlet sector.  (1)-(12) are proportional
to the casimir $C_F$, (13)-(20) to $C_F-C_A/2$,
and (21)-(24) to $C_A$.}
\end{figure}   

Now that we have successfully reduced the number of diagrams we
must calculate from fifty-six (actually, more like seventy-two) 
to the twenty-four shown in Fig. 5, we are ready 
to begin the actual calculation.
Using the fact that $\not\!n\psi_-=\not\!p\psi_+=0$ and that none
of the external momenta have transverse components, one
can reduce every numerator to simple Dirac structures 
times functions of $k_\perp^2$, $p\cdot k$, $\nu$, 
$n\cdot k$, $x$, $y$,
and $x_B$, where $k^\mu$ is the loop-momentum.  
Using techniques outlined in the Appendix, 
one can always cancel propagators until there are only 
three left.  At that point, the integrals can easily 
be done by contour.  Useful integration formulae can be found in the 
Appendix; here, we simply give the result:
\begin{eqnarray}
\label{stns}
S_T^{NS(1)}(\nu,Q^2)&=&{\alpha_s(Q^2)\over2\pi}\sum_q\hat e_q^2
\int\,{dx\,dy\over xy}\,\left\lbrack
M^{NS}\left({x\over x_B},{y\over x_B}\right)-(x_B\rightarrow-x_B)
\right\rbrack
\,K_{qB}(x,y)\;\; ;\\
M^{NS}(x,y)&=&C_F\left\lbrace
{2\over\epsilon}\left({Q^2e^\gamma\over4\pi\mu^2}\right)
^{-\epsilon/2}
\left\lbrack\vphantom{\left(2y(1-x)\over x^2\right)}
-{y\over 1-x}\right.\right.\nonumber\\
&&\left.\phantom{C_F\lbrace}+\left({2y(1-x)\over x^2}-{2\over x}
-{4y\over1-x}\right)\log(1-x)-{2\over y}\log(1-y)
\right\rbrack\nonumber\\
&&\phantom{C_F\lbrace}-{4y\over1-x}+\left({2y\over x^2}+{y-6\over x}
+{2-y\over x-y}-{3y\over 1-x}\right)\log(1-x)\nonumber\\
&&\phantom{C_F\lbrace}+
\left(3-{6\over y}+{y-2\over x-y}\right)\log(1-y)\nonumber\\
&&\phantom{C_F\lbrace}\left.
\vphantom{\left({Q^2e^\gamma\over4\pi\mu^2}\right)^{-\epsilon/2}}
-\left({y(1-x)\over x^2}
-{1\over x}-{2y\over1-x}\right)
\log^2(1-x)+{1\over y}\log^2(1-y)\right\rbrace\label{resultns}\\
&&-{C_A\over2}\left\lbrace
{2\over\epsilon}\left({Q^2e^\gamma\over4\pi\mu^2}\right)
^{-\epsilon/2}
\left\lbrack{2\over y-x}\log(1-x)+{2\over x-y}\log(1-y)
\right\rbrack\right.
\nonumber\\
&&\phantom{-{C_A\over2}\lbrace}+{4\over y-x}\log(1-x)
+\left({4\over x-y}+{2\over 1-x}\right)\log(1-y)\nonumber\\
&&\left.\phantom{-{C_A\over2}\lbrace}
\vphantom{\left({Q^2e^\gamma\over4\pi\mu^2}\right)^{-\epsilon/2}}
-{1\over y-x}\log^2(1-x)
-{1\over x-y}\log^2(1-y)\right\rbrace\nonumber\\
&&+\left(C_F-{C_A\over2}\right)\left\lbrace
{2\over\epsilon}\left({Q^2e^\gamma\over4\pi\mu^2}\right)
^{-\epsilon/2}
{2\over x-y}\left({x+y\over x-y}+{xy\over1-x}\right)
\log(1-(x-y))\right.\nonumber\\
&&\phantom{\left(C_F-{C_A\over2}\right)\lbrace}
+{2\over x-y}\left(2\,{x+y\over x-y}+{y(3x-1)\over1-x}
\right)\log(1-(x-y))
\nonumber\\
&&\left.
\vphantom{\left({Q^2e^\gamma\over4\pi\mu^2}\right)^{-\epsilon/2}}
\phantom{\left(C_F-{C_A\over2}\right)\lbrace}
-{1\over x-y}\left({x+y\over x-y}+{xy\over 1-x}\right)
\log^2(1-(x-y))\right\rbrace\;\; ,\nonumber
\end{eqnarray}
where $\hat e_q^2=e_q^2-\langle e_q^2\rangle$ is the nonsinglet
part of the quark charge.  $\langle e_q^2\rangle=\sum_q e_q^2/n_f$
is the average squared quark charge for $n_f$ quarks.
We have taken $\mu$ as a reference scale for our
regularization procedure, and $\gamma=0.57721\ldots$ is the Euler
constant.
One can check explicitly that $x-y$ divides 
the $C_A$ part of this expression, as required by
gauge invariance.

Interpreting the divergent part of this result
as the renormalization of our distribution,
we find that 
\begin{eqnarray}
&&\int\,dx\left({1\over x-x_B}-{1\over x+x_B}\right)\,
Q^2\,{d\over dQ^2}
q^{NS}_T(x,Q^2)={\alpha_s(Q^2)\over2\pi}\,\int\,{dx \,dy\over xy}\nonumber\\
&&\qquad\times\left\lbrace
C_F\left\lbrack-{y\over x_B-x}+\left({2y(x_B-x)\over x^2}
-2{x_B\over x}-{4y\over x_B-x}\right)\log\left(1-{x\over x_B}\right)
\right.\right.\nonumber\\
&&\left.\qquad\phantom{\times\lbrace C_F\lbrack
}\qquad\vphantom{\left({2y(x_B-x)\over x^2}
-2{x_B\over x}-{4y\over x_B-x}\right)}
-2{x_B\over y}\log\left(1-{y\over x_B}\right)
\right\rbrack\nonumber\\
&&\qquad\phantom{\times\lbrace}
\label{nsev}
-{C_A\over2}\left\lbrack 2{x_B\over y-x}\log\left(1-{x\over x_B}\right)
+2{x_B\over x-y}\log\left(1-{y\over x_B}\right)\right\rbrack\\
&&\qquad\qquad\phantom{\times\lbrace}
+\left(C_F-{C_A\over2}\right)\,2{x_B\over x-y}\left(
{x+y\over x-y}+{xy\over x_B(x_B-x)}\right)
\log\left(1-{x-y\over x_B}\right)\nonumber\\
&&\left.\qquad\qquad\qquad\phantom{\times\lbrace}
\vphantom{{2y(1-x)\over x-r}}-(x_B\rightarrow-x_B)\right\rbrace
K_{NS}(x,y,Q^2)\;\; .\nonumber
\end{eqnarray}
From this result, one can immediately see that the evolution
of $q^{NS}_T(x,Q^2)$ is not self-contained.  This is one of the 
main reasons Eq.~(\ref{mislead}) is misleading.  
The distribution $q^{NS}_T(x,Q^2)$ makes sense by itself only
for one value of $Q^2$.  Once that value is changed, one requires
input from the full function $K_{NS}(x,y,Q^2)$ to define it.

At this point, we can take the imaginary part of Eq.~(\ref{stns})
as $x_B$ approaches the real axis from below to obtain
the next-to-leading expression for $g^{NS}_T(x_B,Q^2)$.  This
is done in Ref. \cite{ournonsing}.  For now, we turn to the 
singlet sector.

\subsection{One-Loop Corrections in the Singlet Sector}

At this order in $\alpha_s(Q^2)$, gluons can interact with the
photons through a virtual quark loop.  This process generates a 
contribution to $S_T$ from purely gluonic effects.  
Using our power-counting techniques, we see that once 
again we must consider three different correlations : $(g,b)=(2,0),
(3,0)$, and $(1,1)$.  As before, the contributions to 
$(2,0)$ can be obtained from those to $(3,0)$.  However, 
due to the bosonic symmetry of the gauge fields, the 
symmetries associated with this process are quite different.

Let us consider first the contributions to $(1,1)$.
These are represented by a quark loop with two photons,
a transverse gluon, and a longitudinal gluon attached.
Including the bosonic symmetries for both gluons and 
photons, a total of six diagrams contribute to this 
amplitude.  Using the properties of fermionic traces,
we can reduce this to the three contributions shown
in Fig. 6.  An explicit calculation shows that these 
contributions exactly cancel.  Since this is the only place
where, through Eq.~(\ref{badg}), 
one can generate a purely singlet quark contribution,
we see that the singlet quark contribution to $S_T^{(1)}$
is identical to the nonsinglet contribution for each quark.

\begin{figure}[t]
\begin{center}
\epsfig{file=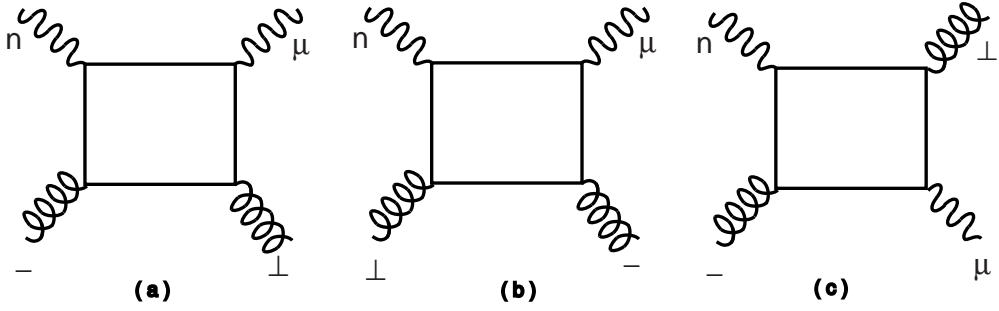,height=4.cm}
\vspace{0.3in}
\end{center}
\caption{Diagrams relevant to $S_T$ at one-loop order in the singlet
sector involving one `bad' gluon field component.
The sum of the three contributions vanishes.}
\end{figure}   

The amplitude associated with $(3,0)$ 
has twenty-four separate diagramatic contributions
once one has taken the three-fold bosonic 
symmetry into account.  Leaving our gluon
indices open, we can reduce this number to 
four.  Once again, the properties of the fermion 
trace allow us to cut this number in half.  
Hence
we are left with only two independent contributions
to our amplitude.  These contributions are 
represented by the diagrams in Fig. 7.  To obtain the 
full amplitude, we simply double the charge-conjugation 
even part of these
diagrams and add the gluon permutations.  A symmetry factor
of $1/6$ must also be applied to associate the amplitude with
our correlation function.  

\begin{figure}[t]
\begin{center}
\epsfig{file=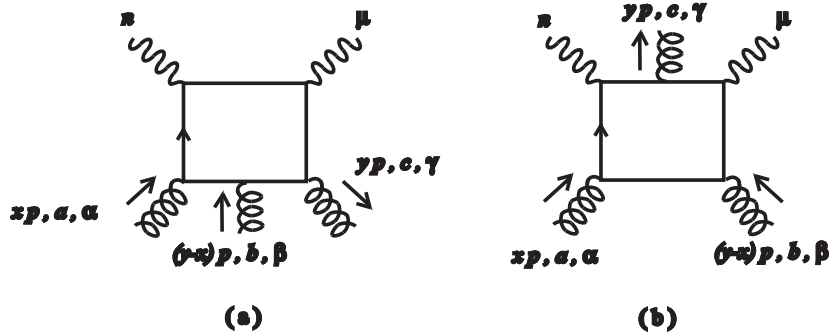,height=4.5cm}
\vspace{0.3in}
\end{center}
\caption{The two independent contributions to  
$M^{\mu+\alpha\beta\gamma\;(g)}_{(3,0)}$.}
\end{figure}   

After simplification, the contributions from Figs. 7 $(a)$ and $(b)$ 
have the form
\begin{eqnarray}
&&M^{\mu+\alpha\beta\gamma\;(g)}_{(3,0)a}\nonumber\\
&&\quad=\left({i\over2}T_F\,f^{abc}\right)(-g_B){\langle 
e_q^2\rangle\over\nu}
\left\lbrack A_1(\tilde x,\tilde y)
g_\perp^{\mu\alpha}g_\perp^{\beta\gamma}
+B_1(\tilde x,\tilde y)g_\perp^{\mu\beta}
g_\perp^{\gamma\alpha}+C_1(\tilde x,\tilde y)
g_\perp^{\mu\gamma}g_\perp^{\alpha\beta}
\right\rbrack\;\; ,\\
&&M^{\mu+\alpha\beta\gamma\;(g)}_{(3,0)b}\nonumber\\
&&\quad=\left({i\over2}T_F\,f^{abc}\right)(-g_B){\langle
e_q^2\rangle\over\nu}
\left\lbrack A_2(\tilde x,\tilde y)
g_\perp^{\mu\alpha}g_\perp^{\beta\gamma}
+B_2(\tilde x,\tilde y)g_\perp^{\mu\beta}g_\perp^{\gamma\alpha}
+C_2(\tilde x,\tilde y)g_\perp^{\mu\gamma}g_\perp^{\alpha\beta}
\right\rbrack\;\; ,
\end{eqnarray}
respectively, where $f^{abc}$ are the structure constants
of the group, $T_F=1/2$ is the generator normalization, 
and $\tilde x=x/x_B$, $\tilde y=y/x_B$ 
represent the relative momentum fractions.  
Here, we have ignored the C-odd $d^{abc}$ contribution
which cannot contribute due to Furry's theorem. 
The full amplitude becomes
\begin{eqnarray}
\label{same}
T^{\mu+\;(g)}_{(3,0)}&=&-{1\over6}\,T_F\,{\langle e_q^2\rangle\over\nu}
\,\int\,dx\,dy\left\lbrack
f\left({y-x\over x_B},-{x\over x_B}\right)
g_\perp^{\mu\alpha}g_\perp^{\beta\gamma}\right.\nonumber\\
&&\phantom{-{1\over6}\,T_F\,{\langle e_q^2\rangle\over\nu}
\,\int\,dx\,dy}+f\left(-{y\over x_B},{x-y\over x_B}\right)
g_\perp^{\mu\beta}g_\perp^{\gamma\alpha}\\
&&\phantom{-{1\over6}\,T_F\,{\langle e_q^2\rangle\over\nu}
\,\int\,dx\,dy}\left.
+f\left({x\over x_B},{y\over x_B}\right)
g_\perp^{\mu\gamma}g_\perp^{\alpha\beta}\right\rbrack\,
(\Gamma_{3gB})_{\gamma\beta\alpha}(x,y)\;\; ;\nonumber\\
f(x,y)&=&\phantom{+}A_1(-y,x-y)+A_2(-y,x-y)-A_1(-y,-x)-A_2(-y,-x)\nonumber\\
&&+B_1(y-x,-x)+B_2(y-x,-x)-B_1(x,x-y)-B_2(x,x-y)\\
&&+C_1(x,y)+C_2(x,y)-C_1(y-x,y)-C_2(y-x,y)\;\; ,\nonumber\\
(\Gamma_{3gB})_{\alpha\beta\gamma}(x,y)&=&\int{d\lambda\over2\pi}\,
{d\mu\over2\pi}\,e^{i\lambda x}\,e^{i\mu(y-x)}
\,\left\langle PS\left|(-igf_{abc})A^a_\alpha(0)A^b_\beta(\mu n)
A^c_\gamma(\lambda n)\right|PS\right\rangle\;\; ,
\end{eqnarray} 
where we have multiplied by a factor of two to take
both possible directions of fermion flow into account
and included a factor of six for gluon symmetry.
Due to the bosonic symmetry of the gauge fields,
all of the terms in Eq.~(\ref{same}) are identical; our
result can be rewritten as
\begin{equation}
T^{\mu+\;(g)}=-{1\over2}\,T_F\,{\langle e_q^2\rangle\over\nu}
\int\,dx\,dy\,f
\left({x\over x_B},{y\over x_B}\right)\,
(\Gamma_{3gB})^{\mu\alpha}_{\;\;\alpha}(x,y)\;\; .
\end{equation}
This amplitude must combine with that associated with
$(2,0)$ to make a fully gauge invariant contribution
to $S^{(1)}_T(\nu, Q^2)$. 

\begin{figure}[t]
\begin{center}
\epsfig{file=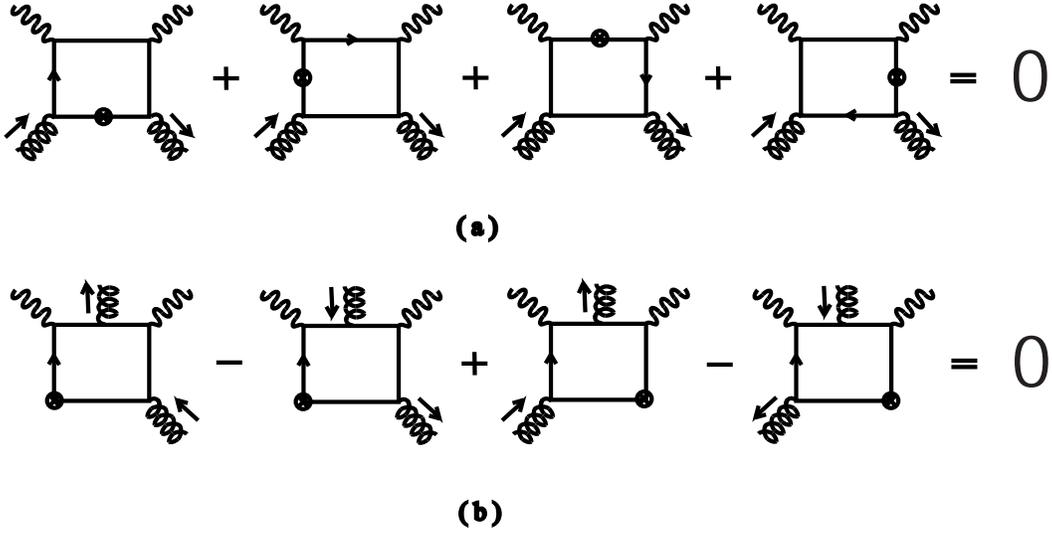,height=7.0cm}
\end{center}
\caption{A diagrammatic representation of the 
consistency relations among loop functions resulting
from our freedom to choose the routing of the 
transverse momentum.}
\end{figure}   

The contributions to $(2,0)$ are calculated according to
the prescription mentioned before.
We send a small amount of transverse momentum into the 
quark loop through the incoming gluon and remove it
through the outgoing.  Internal propagators 
which carry this momentum are then broken successively
by the operator $i\gamma_\perp$.  Since we have a loop, 
we are given a choice of which way to
route the momentum.  Either choice must yield the 
same result, so we immediately have a set of relations
among diagrams.  As before, our new vertices 
differ only trivially from gluon insertions.  Hence the
relations between the $(2,0)$ diagrams translate immediately 
into consistency requirements on the functions
$A_i(x,y)$, $B_i(x,y)$, and $C_i(x,y)$.  
Six relations follow from the two diagramatic
requirements  shown in Fig. 8.  They are
\begin{eqnarray}
\label{relat1}
A_1(0,x)+B_1(x,x)+C_2(x,0)+C_1(x,0)&=&0\;\; ,\\
B_1(0,x)+A_1(x,x)+A_2(x,0)+A_1(x,0)&=&0\;\; ,\\
C_1(0,x)+C_1(x,x)+B_2(x,0)+B_1(x,0)&=&0\;\; ,\\
A_2(x,x)-C_2(-x,-x)+B_2(0,x)-C_2(0,-x)&=&0\;\; ,\\
B_2(x,x)-B_2(-x,-x)+A_2(0,x)-A_2(0,-x)&=&0\;\; ,\\
C_2(x,x)-A_2(-x,-x)+C_2(0,x)-B_2(0,-x)&=&0\;\; .
\label{relat2}
\end{eqnarray}
These relations provide a welcome check of our
calculation, as all of the expressions involved
are extremely complicated.  A similar set of relations
can be derived in the non-singlet sector by allowing the 
transverse momentum to flow through an internal gluon
propagator.  However, in this case the relations 
we would derive are not as useful.

\begin{figure}[t]
\begin{center}
\epsfig{file=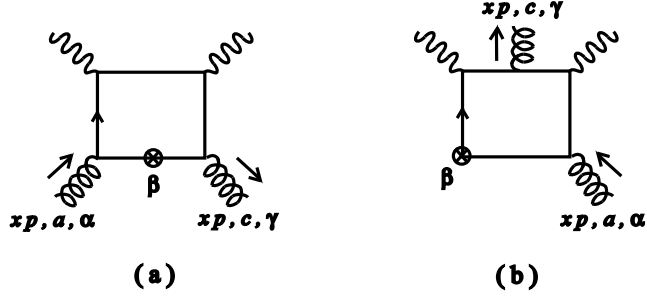,height=4cm}
\end{center}
\caption{The two independent contributions to  
$M^{\mu+\alpha\beta\gamma\;(g)}_{(2,0)}$}
\end{figure}   

Using exactly the same technique that allowed us
to derive the consistency relations, we can 
obtain an expression for the full contribution
to $(2,0)$.  There are eight distinct diagrams, of
which only two are independent.  The diagrams in Fig. 9 $(a)$ and 
$(b)$
have the value
\begin{eqnarray}
M^{\mu+\alpha\beta\gamma\;(g)}_{(2,0)a}=
\left(T_F\delta^{ac}\right)\,{\langle e_q^2\rangle\over\nu}
\left\lbrack
-A_1(\tilde x,\tilde x)g_\perp^{\mu\alpha}g_\perp^{\beta\gamma}
-B_1(\tilde x,\tilde x)g_\perp^{\mu\beta}g_\perp^{\gamma\alpha}
-C_1(\tilde x,\tilde x)g_\perp^{\mu\gamma}
g_\perp^{\alpha\beta}\right\rbrack\;\; ,\\
M^{\mu+\alpha\beta\gamma\;(g)}_{(2,0)b}=
\left(T_F\delta^{ac}\right)\,{\langle e_q^2\rangle\over\nu}
\left\lbrack\phantom{-}
A_2(0,\tilde x)g_\perp^{\mu\beta}g_\perp^{\gamma\alpha}+
B_2(0,\tilde x)g_\perp^{\mu\alpha}g_\perp^{\beta\gamma}+
C_2(0,\tilde x)g_\perp^{\mu\gamma}g_\perp^{\alpha\beta}\right\rbrack\;\; ,
\end{eqnarray}
respectively.
Adding the remaining diagrams and including a 
symmetry factor of 1/2, we find the full contribution
to $(2,0)$ :
\begin{eqnarray}
T^{\mu+\;(g)}_{(2,0)}=T_F\,{\langle e_q^2\rangle\over\nu}
\,\int &dx&\,\left\lbrack
-p_1(-\tilde x)g^{\mu\alpha}_\perp g_\perp^{\beta\gamma}+
p_2(\tilde x)g_\perp^{\mu\beta}g_\perp^{\gamma\alpha}+
p_1(\tilde x)g_\perp^{\mu\gamma}g_\perp^{\alpha\beta}\right\rbrack\,
(\Gamma_{2gB})_{\gamma\beta\alpha}(x)\;\; ;\\
p_1(x)&=&A_1(-x,-x)-C_1(x,x)+C_2(0,x)-B_2(0,-x)\;\; ,\\
p_2(x)&=&B_1(-x,-x)-B_1(x,x)-A_2(0,x)+A_2(0,-x)\;\; ,\\
(\Gamma_{2gB})_{\alpha\beta\gamma}(x)&=&\int\,{d\lambda\over2\pi}
\,e^{i\lambda x}\left\langle PS\left|A^a_\alpha(0)i\partial_\beta
A^a_\gamma(\lambda n)\right|PS\right\rangle\;\; .
\end{eqnarray}
Since our 
amplitude is required to be symmetric under $(x_B\rightarrow-x_B)$,
the 
fact that $p_2(x)$ is an odd function of $x$ implies immediately
that it is zero.  This allows us to write the 
full $(2,0)$ contribution as
\begin{equation}
T^{\mu+\;(g)}_{(2,0)}=2T_F\,{\langle e_q^2\rangle\over\nu}\,
\int dx\,p_1(\tilde x)
(\Gamma_{2gB})^{\mu\alpha}_{\;\;\alpha}(x)\;\; .
\end{equation}

To properly combine with the $(3,0)$ contribution,
it would seem that we must have
\begin{equation}
p_1(x)=-{1\over4}f(x,x)\;\; .
\end{equation}
However, using the relations (\ref{relat1}-\ref{relat2})
one can show that 
\begin{equation}
p_1(x)=-{1\over2}f(x,x)\;\; .
\end{equation}
At first glance, this is disastrous.
It looks very much like we have made a mistake.
On the other hand, we have been very careful to include
all of the appropriate symmetry factors and 
counting.  Going back through the analysis, one
has no choice but to conclude that it is correct.
The answer to our problem lies in a more subtle place.
Looking at $(\Gamma_{3gB})^{\mu\alpha}_{\;\;\alpha}(x,y)$,
we see that it has the peculiar property
\begin{equation}
\int\,dx\,(\Gamma_{3gB})^{\mu\alpha}_{\;\;\alpha}(x,y)=0\;\; .
\end{equation}
Hence there is an ambiguity in our result.  {\it Any}
function $g(y)$ which is independent of $x$ can be added
to $f(x,y)$ to form an equally acceptable amplitude.
The contributing part of $f(x,y)$ is actually the modified
function $f(x,y)-f(0,y)$.  The nature of this ambiguity 
is that it does not affect the $(3,0)$ contribution
to our amplitude.  However, it certainly will affect the 
$(2,0)$ part when one sets $x=y$.  One can easily
show that the modified function satisfies
\begin{equation}
p_1(x)=-{1\over4}\left(f(x,x)-f(0,x)\right)\;\; ,
\end{equation}
as required by gauge invariance.  Defining
the gauge invariant function
\begin{eqnarray}
&&{1\over4}\,xy\,\left\lbrack \Gamma^{\mu\alpha}_{2gB\;\;\alpha}(x,y)
+\Gamma^{\mu\alpha}_{3gB\;\;\alpha}(x,y)\right\rbrack\nonumber\\
&&\qquad=
\int\,{d\lambda\over4\pi}\,{d\mu\over4\pi}\,e^{i\lambda x}
\,e^{i\mu(y-x)}\left\langle PS\left|
F^{+\mu}(0)iD^\perp_\alpha(\mu n)F_\perp^{+\alpha}
(\lambda n)\right|PS\right\rangle
\\
&&\qquad\qquad\qquad\qquad
=i\epsilon^{-+\alpha\mu}MS_\alpha\,K_{gB}(x,y)\;\; ,\nonumber
\end{eqnarray}
and substituting the explicit expression for $f(x,y)$, one 
arrives at the full gluon contribution:
\begin{eqnarray}
\label{singst}
S_T^{(1)g}(\nu,Q^2)&=&
{\alpha_s(Q^2)n_fT_F\over2\pi}\,\langle e_q^2\rangle
\int\,{dx\,dy\over xy(y-x)}\,\left\lbrack
M^g\left(
{x\over x_B},{y\over x_B}\right)\right.\nonumber\\
&&\phantom{{\alpha_s(Q^2)n_fT_F\over2\pi}\,\langle e_q^2\rangle
\int\,{dx\,dy\over xy(y-x)}\lbrack\qquad}\left.\vphantom{
M^g\left(
{x\over x_B},{y\over x_B}\right)}
-(x_B\rightarrow-x_B)\right\rbrack
K_{gB}(x,y)\;\; ;\\
M^g(x,y)&=&{1\over xy}\,\nonumber\\
&&\!\!\!\!\!\!\!\!\!\times\left\lbrace
{2\over\epsilon}\left({Q^2e^\gamma\over4\pi\mu^2}\right)^{-\epsilon/2}
\left\lbrack \vphantom{\left((x+y)(2x+3y)\over x-y\right)}
\left((y-2x)(1-x)+3y(y-x)-6\,{y\over x}\,(y-x)\right)
\log\left(1-x\right)\right.\right.\nonumber\\
&&+\left((4x-3y)(1-y)+4x(y-x)-8\,{x\over y}\,(y-x)\right)
\log\left(1-y\right)\nonumber\\
&&\left.+\left({(x+y)(2x-3y)\over x-y}(1-(x-y))-xy\right)
\log\left(1-(x-y)\right)\right\rbrack
\nonumber\\
&&-3\left(y+2\,{(y-x)^2\over x}\right)(1-x)\log(1-x)\nonumber\\
&&+\left(3(4x-3y)(1-y)
+14x(y-x)-20\,{x\over y}\,(y-x)\right)\log(1-y)\nonumber\\
&&+{6x^2+xy-9y^2\over x-y}(1-(x-y))\log\left(1-(x-y)\right)
\label{singres}\\
&&-{1\over 2}\left((y-2x)(1-x)+3y(y-x)-6\,{y\over x}\,(y-x)\right)
\log^2\left(1-x\right)\nonumber\\
&&-{1\over 2}\left((4x-3y)(1-y)+4x(y-x)-8\,{x\over y}\,(y-x)\right)
\log^2\left(1-y\right)\nonumber\\
&&\left.-{1\over2}\left({(x+y)(2x-3y)\over x-y}(1-(x-y))-xy\right)
\log^2\left(1-(x-y)\right)
\vphantom{{2\over\epsilon}
\left({Q^2e^\gamma\over4\pi\mu^2}\right)^{-\epsilon/2}}
\right\rbrace\;\; ,\nonumber
\end{eqnarray}
which is identical to that in Ref. \cite{oldsing} and 
agrees with the results of Ref. \cite{shoot} after
one has transformed into their basis.

As before, the divergent part of this expression 
generates the evolution of the leading 
distribution.  In this case,
it contributes an extra term to the singlet 
quark distribution, 
\begin{equation}
K_{S}(x,y,Q^2)\equiv \sum_q K_{q}(x,y,Q^2)\;\; ,
\end{equation}
that is not present in the nonsinglet sector :
\begin{eqnarray}
\int\,dx&&\left({1\over x-x_B}-{1\over x+x_B}\right)\,Q^2\,{d\over dQ^2}
q^{S}_T(x,Q^2)=
{\alpha_s(Q^2)n_fT_F\over2\pi}\int\,{dx\,dy\over x^2y^2(y-x)}\nonumber\\
&&\times\left\lbrack\vphantom{x_B\over y}
\left((y-2x)(x_B-x)+3y(y-x)-6\,{x_By\over x}\,(y-x)\right)
\log\left(1-{x\over x_B}\right)\right.\nonumber\\
&&\quad+\left((4x-3y)(x_B-y)+4x(y-x)-8\,{x_Bx\over y}\,(y-x)\right)
\log\left(1-{y\over x_B}\right)\\
&&\quad
\left.+\left({(x+y)(2x-3y)\over x-y}\,(x_B-(x-y))-xy\right)
\log\left(1-{x-y\over x_B}\right)\right.\nonumber\\
&&\qquad\qquad\left.\vphantom{{x_B\over y}}
-(x_B\rightarrow-x_B)\right\rbrack
K_g(x,y,Q^2)+\cdots\;\; ,\nonumber
\end{eqnarray}
where the ellipses denote terms given by Eq.~(\ref{nsev}) above.

At this point, one can easily
obtain an expression for $g_T(x_B,Q^2)$ 
by taking the imaginary part of Eq.~(\ref{singst})
as $x_B$ approaches the positive $x$-axis from below.
This is done explicitly in Ref. \cite{oursing}.

\section{The Operator Product Expansion}

So far, our results have been expressed in terms
of nonlocal parton correlation functions.
While these distributions are easy to interpret, they
do not allow one to exploit the full symmetry of
our theory.  In particular, they 
do not lend themselves to a straightforward
twist analysis. Furthermore, their
evolution equations are unnecessarily complicated.

To obtain an expression with explicit twist separation, we cast
our result in the form of an operator product expansion
of two electromagnetic currents. 
The idea behind this approach is that for large $Q^2$,
the electromagnetic currents in the definition of
$T^{\mu\nu}$ are separated approximately along the 
light-cone direction $n^\mu$.  Such operator products can be
expressed as an infinite sum of local operators in the same
way that a Taylor series can be used to represent a function.
Since local operators can be separated into irreducible
representations of the Lorentz group, they are easier to classify 
and their evolution is somewhat more constrained.

One can obtain the operator product expansion
by expanding our previous results about
$x_B=\infty$ :
\begin{equation}
S_T(\nu,Q^2)=\sum_{n=0}^\infty s_{n}(Q^2)
\left({1\over x_B}\right)^{2n+1}\;\; .
\end{equation}
Although this expansion takes
place in the unphysical region $x_B>1$, 
its terms are related
to moments of the 
physical structure function, $g_T(x_B,Q^2)$ through 
dispersion relations :
\begin{equation}
\int_0^1\,dx\,x^{2n}g_T(x,Q^2)={1\over 4}\,s_{n}(Q^2)\;\; .
\label{disp}
\end{equation}
This shows why the operator product expansion is 
experimentally relevant.  

At leading order in $\alpha_s$, it is quite trivial to 
expand our result. We obtain
\begin{equation}
s_n^{(0)}(Q^2)=2\sum_qe_q^2\int^1_{-1}\,dx\,x^{2n}\,q_T(x,Q^2)\;\; .
\end{equation}
From the definition of $q_T(x,Q^2)$, one can show that
\begin{equation}
\int^1_{-1}\,dx\,x^{2n}\,q_T(x,Q^2)=-{1\over2M^2}
\left\langle PS\left|\overline\psi_qiD_{\mu_1}iD_{\mu_2}\cdots iD_{\mu_{2n}}
\gamma_\alpha\gamma_5\psi_q\right|PS\right\rangle n^{\mu_1}\cdots
n^{\mu_{2n}}MS^\alpha\;\; .
\label{momqt}
\end{equation}
Substituting this result into Eq.~(\ref{disp}) gives us an expression for
the $(2n+1)$-th moment of the physical structure function in terms
of the matrix element of a local operator.  

The different components of the operator above
do not constitute an irreducible representation
of the Lorentz group.  
As such, they are not constrained to have simple
properties as the renormalization scale is changed
or higher-order radiative corrections are taken into account.  
A more useful basis is one which forms irreducible representations
of the Lorentz group.
Such a basis will be useful at arbitrary 
orders in perturbation theory, as the behavior of 
its elements under renormalization 
is restricted by the space-time symmetry.
A standard analysis of the Lorentz group tells us that the
components of the operators
\begin{eqnarray}
\label{tw2q}
_q\theta_{n+1,2}^{\mu_1\cdots\mu_{n+1}}&=&
\overline\psi_qiD^{(\mu_1}\cdots iD^{\mu_{n}}\gamma^{\mu_{n+1})}\gamma_5
\psi_q\;\; ,\\
_q\theta_{n,3}^{\lambda\mu_1\cdots\mu_n}&=&
\overline\psi_qiD^{(\mu_1}\cdots iD^{[\mu_n)}\gamma^{\lambda]}\gamma_5
\psi_q\;\; ,
\end{eqnarray}
where $(\cdots)$ denotes symmetrization of indices and the 
removal of all traces and $[\cdots]$ indicates 
antisymmetrization of indices, form irreducible representations.  
Taking 
\begin{eqnarray} 
\label{me1}
\left\langle PS\left|\,_q
\theta_{n+1,2}^{\mu_1\cdots\mu_{n+1}}\right|PS\right\rangle
&=&2a_n^q(Q^2)\,MS^{(\mu_1}P^{\mu_2}\cdots P^{\mu_{n+1})}\;\; ,\\
\label{me2}
\left\langle PS\left|\,_q\theta^{\lambda\mu_1\cdots\mu_n}_{n,3}
\right| PS\right\rangle
&=&2d_n^q(Q^2)\,MS^{[\lambda}P^{(\mu_1]}\cdots P^{\mu_n)}\;\; ,
\end{eqnarray}
we see that the matrix elements of every component of 
$\theta_{n,3}$ are
kinematically suppressed by at least one 
power of $Q$ in relation to the $+\cdots+$ component of 
$\theta_{n+1,2}$.  This suppression is characterized
by the {\it twist} of the operator, defined as
mass dimension minus spin.  Since $\theta_{n+1,2}$
has spin $n+1$ and $\theta_{n,3}$ has spin $n$, 
the fully symmetric operators
have twist-two and the operators of 
mixed symmetry have twist-three for all values of $n$.
It is generally true that operators of twist $t$ are kinematically
suppressed by at least $(\Lambda_{\rm QCD}/Q)^{t-2}$.  However, 
as exemplified by relations in Eqs.~(\ref{me1},~\ref{me2}),
all of the relevant information can be extracted
at ${\cal O}(\Lambda_{\rm QCD}/Q)^{t-2}$.

In our case, $S_T$ contains contributions
from twist-two and -three.  However, since the twist-two
information can be extracted from the leading
scaling function $g_1(x_B,Q^2)$, only the twist-three
contributions are new.  To see this explicitly, we
transform Eq.~(\ref{momqt}) to the new basis and write
\cite{tree}
\begin{equation}
s_n^{(0)}(Q^2)={2\over 2n+1}\sum_qe_q^2\left\lbrack a^q_{2n}(Q^2)
+2n\,d^{\,q}_{2n}(Q^2)\right\rbrack\;\; .
\end{equation}
An analysis of the leading contributions
to $T^{\mu\nu}$ \cite{twist3} leads one to the moment relations
\begin{equation}
\int_0^1\,dx\,x^n\,g^{(0)}_1(x,Q^2)={1\over 2}\sum_qe_q^2\,a_n^q(Q^2)\;\; ,
\end{equation}
which allows us to isolate the pure twist-three information \cite{wwrel}:
\begin{equation}
\int_0^1\,dx\,x^n\,\left\lbrack
g^{(0)}_T(x,Q^2)-{1\over n+1}\,g^{(0)}_1(x,Q^2)\right\rbrack
={n\over 2(n+1)}\,\sum_qe_q^2\,d_n^q(Q^2)\;\; .
\label{sumrule}
\end{equation}

One of the main purposes of this paper is to present
the one-loop corrections to Eq.~(\ref{sumrule}).  In particular,
it will be interesting to see whether or not the 
twist-three information is isolated by the same combination
of $g_T$ and $g_1$ at higher orders.  A priori, there doesn't
seem to be any reason why it should be.  This statement is really nothing
more than a relation between $g_1$ and the twist-two part
of $g_T$.  It seems to depend on the 
operator (\ref{momqt}) relevant to $g_T$.  In fact, we
will see that this particular combination
of $g_T$ and $g_1$ does isolate pure twist-three effects at one-loop
order.  This fact asserts a kinematic rather than 
dynamic relationship between these two quantities,
as will become clear below.

In order to separate the 
twist-two and -three parts of our one-loop amplitudes, we
must find an unambiguous way to tell the difference
between these contributions.  Since 
one cannot form a mixed-symmetry operator with only 
partial derivatives, $\partial_\perp$ cannot appear 
in the twist-three part of the expansion.  This allows us to
use the partial derivative terms in our distributions 
to identify the coefficients of the twist-two operators.
Completing the operators using the definition in Eq.~(\ref{tw2q}),
we identify the remainder of the expansion 
as purely twist-three.

Before we can systematically express the twist-three 
contribution as a sum of local operators,
we must identify an operator basis in which to express
it.  Unfortunately, the operator $\theta_{n,3}$ is not sufficient
for our needs.  One can immediately see this by looking
at our expression (\ref{resultns}) for the one-loop contribution.
Our expansion will involve general moments of the functions
$K_{q}(x,y,Q^2)$ rather than the simple functions $q_T(x,Q^2)$. 
These moments are directly related to 
matrix elements of operators of the form
\begin{equation}
\overline\psi_q\left(i\partial^+\right)^iiD_\perp^\lambda
\left(i\partial^+\right)^{n-i}\not\!n\gamma_5\psi_q\;\; ,
\end{equation}
but these operators have no definite twist and are
not useful for our purposes.  A useful basis of twist-three
operators in the nonsinglet sector was first identified in
\cite{bkl}.  The operators
\begin{eqnarray}
_qR_{n,j}^{\sigma\mu_1\cdots\mu_n}&=&
\overline\psi_qiD^{(\mu_1}\cdots iD^{\mu_{j-1}}
\left(-igF^{\sigma\mu_j}\right)iD^{\mu_{j+1}}\cdots
iD^{\mu_{n-1}}\gamma^{\mu_n)}\gamma_5\psi\;\; ,\\
_qS_{n,j}^{\sigma\mu_1\cdots\mu_n}&=&
\overline\psi_qiD^{(\mu_1}\cdots iD^{\mu_{j-1}}
\left(g\tilde F^{\sigma\mu_j}\right)iD^{\mu_{j+1}}\cdots
iD^{\mu_{n-1}}\gamma^{\mu_n)}\psi\;\; ;\\
_qW^{\sigma\mu_1\cdots\mu_n}_{n,j}&=&
_qS_{n,j}^{\sigma\mu_1\cdots\mu_n} + 
_qS_{n,n-j}^{\sigma\mu_1\cdots\mu_n} + 
_qR_{n,j}^{\sigma\mu_1\cdots\mu_n} - 
_qR_{n,n-j}^{\sigma\mu_1\cdots\mu_n}\;\; ,
\label{basisq}
\end{eqnarray}
for $j=1$ to $n-1$, 
satisfy 
\begin{equation}
\int\,{dx\,dy\over xy}\,y^j\,x^{n-j}\,(y-x)\,K_q(x,y,Q^2)=
-{1\over8}\,w_{n,j}^q(Q^2)\;\; ,
\label{moments}
\end{equation}
where
\begin{equation}
\left\langle PS\left|\,_qW_{n,j}^{\sigma\mu_1\cdots\mu_n}
\right|PS\right\rangle=2w_{n,j}^q(Q^2)MS^{[\sigma}P^{(\mu_1]}
\cdots P^{\mu_n)}\;\; .
\end{equation}
Note the appearance of the field strength tensor 
in our operator basis.  Since there can be no partial derivative
contributions to the twist-three part of our expansion,
this is required by gauge invariance.  

Expanding $S_T^{NS(1)}$ in Eq.~(\ref{stns}) about $x_B=\infty$ 
and using Eq.~(\ref{moments}), 
we find
\begin{eqnarray}
&&S_T^{NS(1)}(\nu,Q^2)={\alpha_s(Q^2)\over4\pi}\,\sum_q\hat e_q^2
\sum_{n\; {\rm even}=2}^\infty\,\left({1\over x_B}\right)^{n+1}
\left\lbrack \vphantom{-\sum_{j=1}^{n-1}\,c_{n,j}^{3NS}\,w_{n,j}^q(Q^2)}
{2\over n+1}\,c_n^{2NS}\,a_{n}^q(Q^2)\right.\nonumber\\
&&\phantom{S_T^{NS(1)}(\nu,Q^2)={\alpha_s(Q^2)\over4\pi}\,\sum_q\hat e_q^2
\sum_{n\; even=2}^\infty\,\left({1\over x_B}\right)^{n+1}\lbrack}
\left.\qquad
-\sum_{j=1}^{n-1}\,c_{n,j}^{3NS}\,w_{n,j}^q(Q^2)
\right\rbrack\;\; ;\label{stquark}\\
&&c_n^{2NS}=C_F\left\lbrace 4T^1_1(n)-4S_2(n)+\left\lbrack
3+{2\over n+1}+{2\over n+2}\right\rbrack S_1(n)+{8\over n+2}
-{2\over n+1}-9\right\rbrace\;\; ,\\
&&c_{n,j}^{3NS}=(n-j)\left\lbrack {C_F\over2}\left({1\over n+1}
-{2\over n+2}\right)-{1\over2(n+1)}c_n^{2NS}-{C_A\over2}\,{1\over n+2}
\,\left(2+S_1(n+1)\right)\right\rbrack\nonumber\\
&&\phantom{c_{n,j}^{3NS}=}+{C_A\over2}\left\lbrack
S_1(n)-S_1(j)\right\rbrack
+\left(C_F-{C_A\over2}\right)\left\lbrace
\vphantom{\left\lbrack
{1\over n+1}{n-1\choose j-1}
-\sum_{k=j}^{n-1}{k-1\choose j-1}{2+S_1(k)\over k+1}
\right\rbrack}
{1\over n+2}\left( 2+S_1(n+1)\right)-{1\over
n+1}\right.\label{tw3coeff}\\
&&\left.\phantom{c_{n,j}^{3NS}=\left(C_F-{C_A\over2}\right)\lbrace}
+(-1)^j\left\lbrack\vphantom{
+{1\over n+1}{n-1\choose j-1}
-\sum_{k=j}^{n-1}{k-1\choose j-1}{2+S_1(k)\over k+1}
}{1\over n+2}\left\lbrack
{n-1\choose j}-{n-1\choose j-1}
\right\rbrack\left(2+S_1(n+1)\right)\right.\right.
\nonumber\\
&&\left.\left.
\phantom{c_{n,j}^{3NS}=\left(C_F-{C_A\over2}\right)\lbrace}
\qquad
+{1\over n+1}{n-1\choose j-1}
-\sum_{k=j}^{n-1}{k-1\choose j-1}{2+S_1(k)\over k+1}
\right\rbrack\right\rbrace\;\; ,\nonumber
\end{eqnarray}
where we have defined
\begin{eqnarray}
S_j(n)=\sum_{i=1}^n{1\over i^j}\;\; ,\\
T^k_j(n)=\sum_{i=1}^n{S_j(n)\over i^k}\;\; . 
\end{eqnarray}

In the above form, it is easy to see how much the 
result deviates from the naive expectation that
only the twist-three part of $q_T(x,Q^2)$ will
contribute to $S_T$.
The `simple' twist-three operator $\theta_{n,3}$
takes the form 
\begin{equation}
\theta_{n,3}^{\lambda\mu_1\cdots\mu_n}={1\over4n}\sum_{j=1}^{n-1}
(n-j)W^{\lambda\mu_1\cdots\mu_n}_{n,j}
\end{equation}
in the complete basis \cite{bkl}.  In deriving this result, 
the equation of motion has been repeatedly used. Also, like
many of the operator relations in this paper, 
this equation is valid only for forward matrix elements.
Looking at Eq.~(\ref{tw3coeff}), we see that only the 
first term respects the naive expectation.  
Taking the large-$N_c$ limit, which simplifies the 
evolution of $q_T(x,Q^2)$ as we will see below, 
gets rid of most of the complicated $j$-dependence 
of our result, but fails to remove the second term.
All of the $w_{n,j}^q$ enter nontrivially
in the NLO result for $S_T^{NS}$.  This fact will complicate 
precision analysis of $g_T(x, Q^2)$ data immensely.

Since the operators $W_{n,j}$ for different
values of $j$ all have the same mass-dimension and 
transform in the same way under Lorentz rotations, 
they may mix with each other under 
renormalization.  This is 
the origin of all of our difficulties with the evolution of
$q_T(x,Q^2)$.  Since we have no more symmetries to 
constrain the form of our result, a different 
combination of the $W_{n,j}$ will appear at each new
order in $\alpha_s$.  

Expanding Eq.~(\ref{nsev})
about $x_B=\infty$, imposing order-by-order equality, and
transforming to the standard basis,
one arrives at evolution equations for the 
scalar matrix elements of our local operators :
\begin{eqnarray}
Q^2{d\over dQ^2}&&a_n^{NS}(Q^2)={\alpha_s(Q^2)\over2\pi}
C_F\left\lbrack {3\over2}+{1\over (n+1)(n+2)}-2S_1(n+1)\right\rbrack
a_n^{NS}(Q^2)\;\; ,\\
Q^2{d\over dQ^2}&&\sum_{j=1}^{n-1}(n-j)w_{n,j}^{NS}(Q^2)=
{\alpha_s(Q^2)\over2\pi}\,\sum_{j=1}^{n-1}\left\lbrace
\vphantom{+(-1)^j\left({1\over n+2}\left({n-1\choose j}
-{n-1\choose j-1}\right)-\sum_{k=j}^{n-1}{1\over k+1}
{k-1\choose j-1}\right)}
C_F\left\lbrack {3\over2}+{1\over(n+1)(n+2)}-2S_1(n+1)\right\rbrack
\right.\nonumber\\
&&-{C_A\over2}\,{n+1\over n+2}
+\left(C_F-{C_A\over2}\right){n+1\over n-j}\left\lbrack
\vphantom{+(-1)^j\left({1\over n+2}\left({n-1\choose j}
-{n-1\choose j-1}\right)-\sum_{k=j}^{n-1}{1\over k+1}
{k-1\choose j-1}\right)}
{1\over n+2}\right.\\
&&\left.\left.\!\!\!\!\!\!\!\!\!
\!\!+(-1)^j\left({1\over n+2}\left({n-1\choose j}
-{n-1\choose j-1}\right)-\sum_{k=j}^{n-1}{1\over k+1}
{k-1\choose j-1}\right)\right\rbrack\right\rbrace
(n-j)w_{n,j}^{NS}(Q^2)\;\; .\nonumber
\end{eqnarray}
From these equations, it is quite clear that 
diagonal evolution of $_{NS}\theta_{n,3}$ is 
broken only by terms of order $1/N_c^2$ in the 
large-$N_c$ limit.  This fact was first
recognized by Ali, Braun and Hiller \cite{abh} in 1991.  
Its anomalous dimension
in this limit is reproduced by this equation.
The evolution of the $w_{n,j}^{NS}(Q^2)$ was first
obtained by Bukhostov, Kuraev, and Lipatov (BKL)  
\cite{bkl}, and our result agrees with theirs. 
Therefore, we conclude that the amplitude is factorizable 
at this order.
In addition, we reproduce the well-known anomalous dimensions
of the twist-two operators $_{NS}\theta_{n+1,2}$ \cite{dglap}.

In the singlet sector, things become somewhat more
complicated.  Here, we must introduce an 
extra factor of $1/xy$ to obtain gauge-invariance.  
To generate a local gauge-invariant operator product expansion,
it would seem that this factor must be canceled by the 
kernel $M^g(x,y)$.  As argued above, one must {\it define}
$M^g(x,y)$ in such a way that $M^g(0,y)=0$.  Hence, 
the only nontrivial requirement seems to be $M^g(x,0)=0$.
Checking this explicitly, we see that it is not the case.
From this result, one might conclude that a local
gauge-invariant 
operator product expansion does not exist 
for $S_T$ in the singlet channel.  However,
a more careful analysis brings a quite different
result.

The first step in a systematic approach
to this problem lies in identifying and 
separating out the twist-two part of our
amplitude.  Since we know that partial 
derivative terms belong exclusively to this part of the 
amplitude, we can use them as a guide once more.
In analogy with the nonsinglet sector,
we write the twist-two operators
\begin{equation}
_g\theta^{\mu_1\cdots\mu_{n+1}}_{n+1,2}
=F^{\alpha(\mu_1}iD^{\mu_2}\cdots iD^{\mu_n}
i\tilde F^{\mu_{n+1})}_{\;\;\alpha}\;\; ,
\end{equation}
where the dual field strength tensor is defined
by $\tilde F^{\mu\nu}\equiv {1\over2}\epsilon^{\mu\nu\alpha\beta}
F_{\alpha\beta}$.
Taking the $(i+\cdots+)$ component and simplifying,
we obtain
\begin{eqnarray}
_g\theta_{n+1,2}^{i+\cdots+}&=&{2\over n+1}
\epsilon^{ij}
\left\lbrack\vphantom{\sum_{m=0}^{n-1}}
(n+1)iA^j (i\partial^+)^n i\partial^k A^k
+iA^j (i\partial^+)^{n-1}D_\alpha F^{\alpha +}\right.\nonumber\\
&&+
\left.f^{abc}A_a^j(i\partial^+)^{n-1}gA_b^k i\partial^+A_c^k+
\sum_{m=0}^{n-1}f^{abc}A_a^j(i\partial^+)^mgA_b^k(i\partial^+)^{n-m}A_c^k
\right\rbrack\;\; .
\label{twist2g}
\end{eqnarray}
Here, the indices $(i,j,k)$ represent transverse dimensions.
Note that a sign has been taken into account for 
transverse partial derivatives and gauge fields with contravariant
indices.  There are several important features
of this operator.  One is the appearence of the 
operator $D_\alpha F^{\alpha+}$.  Through the 
gluon equations of motion, this term 
transmutes into a quark singlet contribution.
It will generate a difference between the 
behavior of the singlet and nonsinglet twist-three quark
operators under renormalization and their contributions
to the operator product expansion.  At present, 
its most striking feature is that it contains
terms which can be generated by $M^g(x,0)$.  
This suggests that separating the twist-two
contribution, whose gauge invariance and locality are
not in question, from the amplitude {\it before} worrying
about the issues of gauge invariance may lead us to 
the correct understanding of our result.

In order to perform a straightforward separation, it is 
advantageous for us to write the three-gluon part of
our amplitude in such a way that the bosonic 
symmetry is evident.  As we have seen above, 
forgetting this symmetry can easily lead to 
incorrect conclusions.
To this end, we define the local operator 
\begin{equation}
{\cal O}_{n,j}^\mu=igf^{abc}A_a^\mu\left\lbrack
(i\partial^+)^j A_\alpha^b\right\rbrack\left\lbrack
(i\partial^+)^{n-j}A_c^\alpha\right\rbrack
\end{equation}
and re-write our twist-two operator in terms
of it :
\begin{eqnarray}
_g\theta_{n+1,2}^{i+\cdots+}=-{2\over n+1}
\epsilon^{ij}&&\left\lbrace 
\vphantom{-i\sum_{j=1}^{n-1}\left\lbrack{n-1\choose j}
+{n\choose j+1}\right\rbrack{\cal O}_{n,j}}(n+1)iA^j
iD^\perp_\sigma(i\partial^+)^nA_\perp^\sigma
-iA^j(i\partial^+)^{n-1}D_\alpha F^{\alpha+}\right.\nonumber\\
&&\left.-i\sum_{k=1}^{n-1}\left\lbrack{n-1\choose k}
+{n\choose k+1}\right\rbrack{\cal O}_{n,k}^j\right\rbrace\;\; .
\end{eqnarray}
This equation is valid modulo total derivatives.
Note that bosonic symmetry implies ${\cal O}^\mu_{n,j}
=-{\cal O}^\mu_{n,n-j}$.
Our full amplitude takes the form
\begin{eqnarray}
T^{\mu+(1)}_{g}&=&{\alpha_sn_fT_F\over2\pi}\,{\langle e_q^2\rangle\over\nu}
\sum_{n\;\; {\rm even}=2}^{\infty}\left({1\over x_B}\right)^n\nonumber\\
&&\times\left\lbrace\vphantom{-{2\over(n+1)(n+2)}
\left\lbrack S_1(n)+2\right\rbrack
\sum_{j=1}^{n-1}{n\choose j+1}\left\langle
PS\left|{\cal O}^\mu_{n,j}\right|PS\right\rangle}
4\left\lbrack{2\over n+2}-{1\over n+1}\right\rbrack
\left\lbrack S_1(n)+1\right\rbrack\left\langle
PS\left|A^\mu iD^\perp_\sigma(i\partial^+)^nA_\perp^\sigma
\right|PS\right\rangle\right.\nonumber\\
&&\left.\phantom{\times\lbrace}
-{2\over(n+1)(n+2)}\left\lbrack S_1(n)+2\right\rbrack
\sum_{j=1}^{n-1}{n\choose j+1}\left\langle
PS\left|{\cal O}^\mu_{n,j}\right|PS\right\rangle\right\rbrace\;\; ,
\label{stglue}
\end{eqnarray}
where the divergent part has been ignored.  It
can be restored via the substitution
\begin{equation}
S_1(n)\rightarrow {2\over\epsilon}\left({Q^2e^\gamma\over4\pi\mu^2}\right)
^{-\epsilon/2}+S_1(n)\;\; .
\end{equation}

In this form, it is easy to separate the 
twist-two and -three parts of the amplitude.  We
have only to identify an appropriate basis of 
twist-three gluonic operators to write the final result.
An appropriate basis takes
\begin{eqnarray}
_gS_{n,j}^{\lambda\mu_1\cdots\mu_n}&=&
F^{\alpha(\mu_1}iD^{\mu_2}\cdots iD^{\mu_{j-1}}
ig\tilde F^{\lambda\mu_j}iD^{\mu_{j+1}}\cdots iD^{\mu_{n-1}}
F^{\mu_n)}_{\;\;\alpha}\;\; ,\\
_gW_{n,j}^{\lambda\mu_1\cdots\mu_n}&=&
_gS_{n,j}^{\lambda\mu_1\cdots\mu_n}+\,
_gS_{n,n-j+1}^{\lambda\mu_1\cdots\mu_n}
\end{eqnarray}
for $j=2$ to $n-1$.  The analogue of $_qR_{n,j}$
is not independent in this case since 
\begin{equation}
F^{\alpha+}(i\partial^+)^i\tilde F^{\lambda+}(i\partial^+)^j
F^{+}_{\;\;\alpha}=F^{\alpha+}(i\partial^+)^iF^{+}_{\;\;\alpha}
(i\partial^+)^j\tilde F^{\lambda+}+
F^{\lambda+}(i\partial^+)^i\tilde F^{\alpha+}(i\partial^+)^j
F^{+}_{\;\;\alpha}\;\; .
\end{equation}
Note that our basis differs from that found in Ref.~\cite{bkl}.
Through repeated partial integration and 
binomial coefficient sums, one can show
that the relation
\begin{equation}
\sum_{j=1}^{n-1}c_{n,j}{\cal O}^\mu_{n,j}
=i\epsilon^{\mu+-\beta}g_{\beta\lambda}\sum_{j=2}^{n-1}
{1\over2}\sum_{k=j}^{n-1}(-1)^{k+1}c_{n,k}\,
_gW^{\lambda+\cdots+}_{n,j}
\end{equation}
is valid for forward matrix elements.
It is this expression that establishes the gauge invariance
of our result.  Defining scalar matrix elements in
exact analogy to the nonsinglet sector, we find the 
following result for the singlet part of $S_T^{(1)}$ :
\begin{eqnarray}
&&S_T^{S(1)}(\nu,Q^2)={\alpha_s(Q^2)n_fT_F\over2\pi}\,\langle e_q^2\rangle
\sum_{n\; {\rm even}=2}^\infty\,\left({1\over x_B}\right)^{n+1}\nonumber\\
&&\qquad\times\left\lbrace {2n\over(n+1)^2(n+2)}\left(
S_1(n)+1\right)\left\lbrack
-2a_n^g(Q^2)+\sum_{j=1}^{n-1}{n-2\choose j-1}
(-1)^{j+1}w_{n,j}^S(Q^2)\right\rbrack\right.\nonumber\\
&&\qquad\qquad\qquad\left.+{1\over2}\,{1\over(n+1)(n+2)}\sum_{j=2}^{n-1}
c_{n,j}^gw^g_{n,j}(Q^2)\right\rbrace\label{thisone}\\
&&\qquad\qquad+{\alpha_s(Q^2)\over4\pi}\,\langle e_q^2\rangle
\sum_{n\; {\rm even}=2}^\infty\,\left({1\over x_B}\right)^{n+1}
\left\lbrack {2\over n+1}\,c_n^{2NS}\,a_{n}^S(Q^2)
-\sum_{j=1}^{n-1}\,c_{n,j}^{3NS}\,w_{n,j}^S(Q^2)
\right\rbrack\;\; ;\nonumber\\
&&c_{n,j}^g=n-1-{n^2+1\over n+1}\left(S_1(n)+1\right)\nonumber\\
&&\qquad\qquad+2(-1)^j{n-1\choose j}\left\lbrack
1-{(n-1)^2+2nj\over n^2-1}\left(S_1(n)+1\right)\right\rbrack\;\; .
\end{eqnarray}
Here, the scalar matrix elements of 
singlet quark operators are defined via
\begin{eqnarray}
a_n^S(Q^2)&=&\sum_q a_n^q(Q^2)\;\; ,\\
w_{n,j}^S(Q^2)&=&\sum_q w_{n,j}^q(Q^2)\;\; .
\end{eqnarray}
This result was presented in a different 
operator basis in \cite{shoot} for $n=2,4,6$.    

The evolution in the singlet sector can also be 
deduced by our results.  One obtains
\begin{eqnarray}
Q^2{d\over dQ^2}&&a_n^{S}(Q^2)={\alpha_s(Q^2)\over2\pi}
C_F\left\lbrack {3\over2}+{1\over (n+1)(n+2)}-2S_1(n+1)\right\rbrack
a_n^{S}(Q^2)\nonumber\\
&&\phantom{a_n^{S}(Q^2)}+{\alpha_s(Q^2)\over2\pi}\,2n_fT_F\left({2\over n+2}
-{1\over n+1}\right)a_n^g(Q^2)\;\; ,\\
Q^2{d\over dQ^2}&&\sum_{j=1}^{n-1}(n-j)w_{n,j}^{S}(Q^2)=
{\alpha_s(Q^2)\over2\pi}\,\sum_{j=1}^{n-1}\left\lbrace
\vphantom{\left\lbrack\left({n\choose j}\right)\right\rbrack}
C_F\left\lbrack {3\over2}+{1\over(n+1)(n+2)}-2S_1(n+1)\right\rbrack
\right.\nonumber\\
&&\qquad+4n_fT_F{(-1)^j\over n-j}\,{n\over(n+1)(n+2)}\,
{n-2\choose j-1}-
{C_A\over2}\,{n+1\over n+2}\nonumber\\
&&\qquad+\left(C_F-{C_A\over2}\right){n+1\over n-j}\left\lbrack
\vphantom{\left({1\over n+2}\left({n-1\choose j}
-{n-1\choose j-1}\right)-\sum_{k=j}^{n-1}{1\over k+1}
{k-1\choose j-1}\right)}
{1\over n+2}\right.\\
&&\left.\left.+(-1)^j\left({1\over n+2}\left({n-1\choose j}
-{n-1\choose j-1}\right)-\sum_{k=j}^{n-1}{1\over k+1}
{k-1\choose j-1}\right)\right\rbrack\right\rbrace
(n-j)w_{n,j}^{S}(Q^2)\nonumber\\
&&+{\alpha_s(Q^2)\over4\pi}\,2n_fT_F\sum_{j=2}^{n-1}
{1\over n+2}\left\lbrack {n^2+1\over n+1}+2(-1)^j
{n-1\choose j}\,{(n-1)^2+2jn\over n^2-1}\right\rbrack
w^g_{n,j}(Q^2)\;\; .\nonumber
\end{eqnarray}
Here, also, one reproduces the well-known twist-two result
\cite{dglap}.
Use of the equation of motion in the 
gluonic twist-two operator has generated a term
which destroys the autonomous evolution of
$_S\theta_{n,3}$ in the large-$N_c$ limit.
This behavior can be predicted simply by looking 
at the diagrams which contribute to evolution in this
sector, c.f. \cite{ourlargen}.  These results were
first obtained by BKL \cite{bkl}.  Translating from
their basis to ours, we find that there is some
disagreement between our results and theirs. 

We are now in a position to present the 
full NLO corrections to Eq.~(\ref{sumrule}).
Using the well-known result \cite{wc,kodaira,g1sum}
\begin{eqnarray}
\int_0^1 x^n g_1(x,Q^2)&=&
{1\over 2}\left\lbrack 1+{\alpha_s(Q^2)\over4\pi}\,c^{2NS}_n
\right\rbrack\sum_qe_q^2a_n^q(Q^2)\nonumber\\
&&-{\alpha_s(Q^2)n_fT_F\over 2\pi}
\,\langle e_q^2\rangle\,{n\over(n+1)(n+2)}\,\left(S_1(n)+1\right)
a_n^g(Q^2)\;\; ,
\end{eqnarray}
one finds
\begin{eqnarray}
\int_0^1\,dx\,x^n&&\left(g_T(x,Q^2)-{1\over n+1}\,g_1(x,Q^2)\right)
={1\over8(n+1)}\,\sum_qe_q^2\nonumber\\
&&\times\left\lbrace\sum_{j=1}^{n-1}\left\lbrack
1-{\alpha_s(Q^2)\over2\pi}\,{n+1\over n-j}\,c_{n,j}^{3NS}\right\rbrack
(n-j)w^q_{n,j}(Q^2)\right.\nonumber\\
&&\phantom{\times\lbrace}+{\alpha_s(Q^2)T_F\over\pi}\,
{2n\over(n+1)(n+2)}\left( S_1(n)+1\right)
\sum_{j=1}^{n-1}\,(-1)^{j-1}{n-2\choose j-1}w_{n,j}^S(Q^2)\\
&&\left.\phantom{\times\lbrace}+{\alpha_s(Q^2)T_F\over2\pi}
\,{1\over n+2}\sum_{j=2}^{n-1}c_{n,j}^g 
w^g_{n,j}(Q^2)\right\rbrace
+{\cal O}\left(\Lambda_{\rm QCD}/Q,\alpha_s^2(Q^2)\right)\;\; .\nonumber
\end{eqnarray}
As promised, the relation between $g_1$ and the twist-two 
part of $g_T$ is unaffected at this order.  

We can gain insight into this phenomenon 
by stripping away the external states 
implicit in Eqns.(\ref{stquark},\ref{thisone})
and writing the operator product expansion in its
full glory :
\begin{eqnarray}
i\int\,d^4z&&\;e^{iq\cdot z}TJ^{[\mu}(z) J^{\nu]}(0)=
-i\epsilon^{\mu\nu\alpha\beta}q_\alpha g_{\beta\lambda}\,{2\over Q^2}
\sum_{n\; {\rm even}=2}^\infty {(2q_{\mu_1})\cdots(2q_{\mu_n})\over (Q^2)^n}
\nonumber\\
&&\times\left\lbrace 
\vphantom{+{\alpha_s(Q^2)n_fT_F\over4\pi}
\,\langle e_q^2\rangle\,{1\over (n+1)(n+2)}\sum_{j=2}^{n-1}
c^g_{n,j}\,_gW^{\lambda\mu_1\cdots\mu_n}_{n,j}}
\sum_qe_q^2\left\lbrack 1+{\alpha_s(Q^2)\over4\pi}
c_n^{2NS}\right\rbrack\;_q\theta_{n+1,2}^{\lambda\mu_1\cdots\mu_n}
\right.\nonumber\\
&&\phantom{\times\lbrace}
-{\alpha_s(Q^2)n_fT_F\over\pi}\,\langle e_q^2\rangle\,{n\over (n+1)(n+2)}
\left\lbrack S_1(n)+1\right\rbrack\;
_g\theta^{\lambda\mu_1\cdots\mu_n}_{n+1,2}\nonumber\\
&&\phantom{\times\lbrace}+\sum_qe_q^2
{1\over 2(n+1)}\sum_{j=1}^{n-1}\left\lbrack
1-{\alpha_s(Q^2)\over2\pi}\,{n+1\over n-j}\,c_{n,j}^{3NS}\right\rbrack
(n-j)\;_qW^{\lambda\mu_1\cdots\mu_n}_{n,j}\\
&&\phantom{\times\lbrace}
+{\alpha_s(Q^2)n_fT_F\over\pi}
\,\langle e_q^2\rangle\,{n\over(n+1)^2(n+2)}\left(S_1(n)+1\right)
\sum_{j=1}^{n-1}(-1)^{j-1}{n-2\choose j-1}\;
_SW^{\lambda\mu_1\cdots\mu_n}_{n,j}\nonumber\\
&&\left.+{\alpha_s(Q^2)n_fT_F\over4\pi}
\,\langle e_q^2\rangle\,{1\over (n+1)(n+2)}\sum_{j=2}^{n-1}
c^g_{n,j}\;_gW^{\lambda\mu_1\cdots\mu_n}_{n,j}\right\rbrace
+{\cal O}\left(\Lambda^2_{\rm QCD}/Q^2,
\alpha_s^2(Q^2)\right)\;\; .\nonumber
\end{eqnarray}
In this form, it is easy to see that the relationship
between $g_1(x,Q^2)$ and the twist-two part of $g_T(x,Q^2)$ 
is kinematical in nature.  The operators involved
are identical, so one expects their dynamical coefficients in
this expansion to be identical.  
Stated another way, the twist-two contribution
to this expansion is independent of the 
kinematics associated with the matrix elements
we choose to take.  From this point of view,
one expects this behavior at all orders of perturbation theory.
Inverting the Mellin transform, one obtains the functional
relationship \cite{wwrel}
\begin{equation}
g_1(x,Q^2)=g_T^{\rm tw-2}(x,Q^2)x\delta(x-1)
-x{d\over dx}\,g^{\rm tw-2}_T(x,Q^2)\;\; ,
\end{equation}
which must be satisfied if the 
operator product expansion is truly well-defined.

\section{Functional Twist Separation}

Although we have managed twist separation
for the individual moments of the 
structure function, it is natural to ask
if this separation can be done in term of the
correlation functions themselves.  
This procedure would allow us to express
$g_T^{\rm tw-2}(x_B,Q^2)$ in terms of twist-two
distributions, for example.
The most straightforward way of achieving this
involves a simple resummation of the separate contributions
to Eqs.~(\ref{stquark},\ref{thisone}).  However, a 
quick glance at their form is more than enough to 
motivate us to find another way.

The twist associated with a certain 
functional form is only well-defined 
after expansion since, as mentioned above, 
nonlocal operators do not form irreducible representations
of the Lorentz group.  However, it is quite easy
to construct functions whose expansions display the symmetries of
a given twist.  For example, the form of the 
non-singlet twist-two operator (\ref{tw2q}) implies
that for any function $f(x)$ analytic in a region
about the origin, the function  
\begin{equation}
T_2(x,y)=f(x)+{xf(x)\over x-y}-{yf(y)\over x-y}
\end{equation}
will lead exclusively to twist-two local operators 
upon convolution with $K_q(x,y,Q^2)$.
Similarly, it is obvious from our operator
basis (\ref{basisq}) that for any function
$f(x,y)$ that is regular at $x=y$, the function
\begin{equation}
T_3(x,y)=(y-x)f(x,y)
\end{equation}
will lead exclusively to twist-three local operators
upon convolution with $K_q(x,y,Q^2)$.  
This fact is especially useful since it allows us
to isolate the part of our amplitude which does not 
vanish as $x\rightarrow y$ into a function 
only of $x$, with corrections that contribute 
directly to the twist-three kernel :
\begin{equation}
f(x,y)=\lim_{y\rightarrow x} f(x,y)+\left\lbrack f(x,y)
-\lim_{y\rightarrow x} f(x,y)\right\rbrack\;\; .
\end{equation}
Now, we need only construct a way to separate the twist-two
and -three contributions to a function only of $x$.

In the non-singlet sector, this separation is quite 
straightforward.  Since convoluting pure powers
of $x$ with $K_q(x,y,Q^2)$ generates moments of $q_T(x,Q^2)$,
one finds that the expression
\begin{equation}
x^{n-1}={1\over n+1}\left\lbrack
x^{n-1}+\sum_{i=0}^{n-1}\,x^iy^{n-i-1}\right\rbrack
+{2n\over n+1}\,{1\over 2n}\left\lbrack nx^{n-1}-\sum_{i=0}^{n-1}
\,x^iy^{n-i-1}\right\rbrack
\end{equation}
separates the twist-two and -three components
of $x^{n-1}$.  This can be translated into a 
functional relationship quite easily :
\begin{eqnarray}
f(x)&=&\left\lbrack g(x)+{xg(x)\over x-y}-{yg(y)\over x-y}\right\rbrack
\nonumber\\
&&+\left\lbrack {d\over dx}\,xg(x)-{xg(x)\over x-y}+
{yg(y)\over x-y}\right\rbrack\;\; ;\\
g(x)&=&{1\over x^2}\int_0^x\,dx'\,x'\,f(x')\;\; .
\end{eqnarray}
Hence for any kernel $M^{NS}(x,y)$ in the non-singlet sector,
the twist-two and -three parts are given by
\begin{eqnarray}
&&m^{NS}(x)+{x\,m^{NS}(x)\over x-y}-{y\,m^{NS}(y)\over x-y}\;\; ,\\
&&{d\over dx}\,x\,m^{NS}(x)-{x\,m^{NS}(x)\over x-y}+{y\,m^{NS}(y)\over x-y}
+\left\lbrack M^{NS}(x,y)-\lim_{y\rightarrow x}M^{NS}(x,y)
\right\rbrack\;\; ,
\end{eqnarray}
respectively, where 
\begin{equation}
m^{NS}(x)\equiv{1\over x^2}\int_0^x\,dx'\,x'\lim_{y\rightarrow x'}
M^{NS}(x',y)\;\; .
\end{equation}

While conceptually simple, this functional separation
is practically quite complicated.  As an example, 
we examine the leading order.  Here, 
\begin{equation}
m^{NS}(x)=-{2\over x^2}\int_0^x\,{dx'\over x'-1}=-{2\over x^2}
\log(1-x)\;\; ,
\end{equation}
which leads to
\begin{eqnarray}
S_T^{NS(0)\rm tw-2}(\nu,Q^2)&=&-\sum_qe_q^2\int\,dxdy\,\left\lbrack
\phantom{+}
{2\over x^2}\log\left(1-{x\over x_B}\right)\right.\nonumber\\
&&\phantom{-\sum_qe_q^2\int\,dxdy\,\lbrack}
+{2\over x(x-y)}\,\log\left(1-{x\over x_B}\right)\\
&&\left.\phantom{-\sum_qe_q^2\int\,dxdy\,\lbrack}
-{2\over y(x-y)}\,\log\left(1-{y\over x_B}\right)\right\rbrack
K_q(x,y,Q^2)\;\; ,\nonumber\\
S_T^{NS(0)\rm tw-3}(\nu,Q^2)&=&-\sum_qe_q^2\int\,dxdy\,\left\lbrack
\phantom{+}
{2\over x(x-x_B)}-{2\over x^2}\log\left(1-{x\over x_B}\right)
\right.\nonumber\\
&&\phantom{-\sum_qe_q^2\int\,dxdy\,\lbrack}
-{2\over x(x-y)}\,\log\left(1-{x\over x_B}\right)\\
&&\phantom{-\sum_qe_q^2\int\,dxdy\,\lbrack}\left.
+{2\over y(x-y)}\,\log\left(1-{y\over x_B}\right)\right\rbrack
K_q(x,y,Q^2)\;\; .\nonumber
\end{eqnarray}
Obviously, one can separate the kernel (\ref{resultns}) in a similar
way.  However, the result of such a manipulation is extremely
complicated and not very illuminating; it will not be presented here.

The advantage of performing this separation
is that we can explicitly replace $K_q(x,y,Q^2)$
with the simple distribution
\begin{equation}
\Delta q(x,Q^2)=\int\,{d\lambda\over 4\pi}
e^{i\lambda x}\left\langle PS\left|
\overline\psi_q(0)\not\!n\gamma_5\psi_q(\lambda n)\right|PS\right\rangle
\end{equation}
in our expression for $S_T^{NS(0)\rm tw-2}(\nu,Q^2)$.
Since 
\begin{eqnarray}
\int\,dxdy\left\lbrack x^{n-1}+\sum_{i=0}^{n-1}x^iy^{n-1-i}\right\rbrack
K_q(x,y,Q^2)&=&-{n+1\over 4M}\,S_\alpha\left\langle
PS\left|\,_q\theta_{n+1,2}^{\alpha+\cdots+}\right|PS\right\rangle\nonumber\\
&&={1\over 2}\,a_n^q(Q^2)={1\over 2}\int\,dx\,x^n\Delta q(x,Q^2)\;\; ,
\end{eqnarray}
one can write 
\begin{equation}
S_T^{NS(0)\rm tw-2}(\nu,Q^2)=-\sum_qe_q^2\int {dx\over x}\,\log\left(
1-{x\over x_B}\right)\Delta q(x,Q^2)\;\; .
\end{equation}
More generally, for any kernel $M^{NS}(x,y)$ in the 
non-singlet sector, the twist-two part can be written 
\begin{equation}
{1\over2}\int dx\,x\,m^{NS}(x)\Delta q(x)\;\; ,
\end{equation}
where $m^{NS}(x)$ is given above.

In the singlet sector, the separation is spoiled by the presence
of $D_\alpha F^{\alpha+}$ in Eq.~(\ref{twist2g}).  This
makes it impossible for us to express the twist-two
gluon operator in terms of $K_g(x,y,Q^2)$.  One can 
circumnavigate this problem by  
introducing the singlet quark distribution.  Examining the 
twist-two gluon operator (\ref{twist2g}), we see that
\begin{eqnarray}
&&\int{dx\,dy\over xy}\left\lbrack
x\,y^{n-1}+\sum_{i=0}^{n-1}\,y^i\,x^{n-i}\right\rbrack\,K_g(x,y,Q^2)
\nonumber\\
&&\qquad+\int\,dx\,dy\,(y-x)^{n-1}\,K_S(x,y,Q^2)={1\over4}\,a_g^n(Q^2)
\end{eqnarray}
for $n$ even.  In analogy with the non-singlet sector,
we use this to separate the amplitude
\begin{equation}
S_T^g(\nu,Q^2)=\int\,{dx\,dy\over xy}\,{1\over y-x}\,M^g\left(
{x\over x_B},{y\over x_B}\right)K_g(x,y,Q^2)
\end{equation}
into its twist-two,
\begin{equation}
S_T^{g\; \rm tw-2}(\nu,Q^2)={1\over 4}\,\int {dx\over x}\,m^g
\left({x\over x_B}\right)\Delta g(x)\;\; ,
\end{equation}
and -three,
\begin{eqnarray}
S_T^{g\;\rm tw-3}(\nu,Q^2)&=&\phantom{+}\int\,{dx\,dy\over xy}\left\lbrack
{1\over y-x}M^g\left({x\over x_B},{y\over x_B}\right)
-{x\over y^2}\,m^g\left({y\over x_B}\right)\right.\nonumber\\
&&\phantom{+\int\,{dx\,dy\over xy}\lbrack}\left.
-{x\over y(y-x)}\,m^g\left({y\over x}\right)
+{1\over y-x}\,m^g\left({x\over x_B}\right)\right\rbrack
K_g(x,y,Q^2)\\
&&-\int\,{dx\,dy\over (y-x)^2}\,m^g\left({y-x\over x_B}\right)
K_S(x,y,Q^2)\;\; ,\nonumber
\end{eqnarray}
parts.  Here, we have defined
\begin{equation}
m^g(x)=\int_0^x\,dx'\lim_{y\rightarrow x'}
{1\over y-x'}\,M^g(x',y)\;\; ,
\end{equation}
and introduced
\begin{equation}
\Delta g(x)={i\over x}\,\int{d\lambda\over 4\pi}\,e^{i\lambda x}\,
\left\langle PS\left|
F^{\alpha+}(0)\tilde 
F^{+}_{\;\;\alpha}(\lambda n)\right|PS\right\rangle\;\; .
\end{equation}
Once again, explicit expressions for these 
separate kernels could obviously be obtained 
from (\ref{singres}).  However, due to the form of
that result, such manipulations are better left for numerical
analysis.

\section{Conclusion}

We have presented a systematic way to obtain
radiative corrections to the power-suppressed terms in 
inclusive deep-inelastic scattering.  This method is quite 
general and can be applied to other factorizable processes 
such as deeply-virtual Compton scattering and 
Drell-Yan scattering with few changes.

Using this method, we have calculated the 
next-to-leading order corrections 
to the Compton amplitude 
$S_T(\nu,Q^2)$.  This result allows us to 
derive an expression for the 
operator product expansion of 
two electromagnetic currents separated
along the light-cone valid to twist-three
and ${\cal O}(\alpha_s)$.  
Our results show that the 
Wandzura-Wilczek relation between $g_1$ and the 
twist-two part of $g_T$ is valid to 
this order and the form of the expansion
indicates its validity to 
all orders in perturbation theory.
The large-$N_c$ 
simplification of the evolution of 
$q^{NS}_T(x,Q^2)$ does not extend to 
its coefficient function at next-to-leading order.
The general distributions $K_{NS}(x,y,Q^2)$
enter nontrivially even in this limit.

It would be interesting to apply
our method to higher twist processes.
In particular, ignoring interactions in the
nuclear wavefunction, certain nuclear twist-four
matrix elements can be factorized into the product
of two nucleonic twist-two distributions.  
In view of recent data from RHIC, corrections to these
processes may allow us to clearly determine the 
changes in parton distributions induced by the nuclear
medium.  In addition, radiative corrections to the 
nucleon's unpolarized longitudinal structure function
$F_L(x,Q^2)$ can be analyzed with this technique.

\acknowledgements
The authors would like to thank A. Belitsky, 
W. Lu and X. Song for collaboration at early stages 
of this work. The work is supported in part by
the Director, Office of Science, Office of High Energy and Nuclear
Physics, and by the Office of Basic Energy Sciences, 
Division of Nuclear Sciences of  
the U.S.~Department of Energy under grant nos. 
DE-FG02-93ER-40762 and DE-AC03-76SF-00098.

\begin{appendix}
\section{}
This appendix is meant to provide some 
formulae we have found useful during this
calculation.  The Feynman integrals 
one comes across here are either of the form
\begin{equation}
\int{d^dk\over(2\pi)^d}\,{(2p\cdot k)^{n-i-1}(k_\perp^2)^{i+m-1}
\prod_{\ell=1}^j(k\cdot n+a_\ell)\over
k^2(k+y_1p)^2\cdots(k+y_np)^2(k+x_1p+q)^2\cdots
(k+x_mp+q)^2}\phantom{\;\; ,}
\label{int1}
\end{equation}
or
\begin{equation}
\int{d^dk\over(2\pi)^d}\,{(2p\cdot k)^{n-i}(k_\perp^2)^{i+m-1}
\prod_{\ell=1}^j(k\cdot n+a_\ell)\over
k^2(k+y_1p)^2\cdots(k+y_np)^2(k+x_1p+q)^2\cdots
(k+x_mp+q)^2}\;\; ,
\label{int2}
\end{equation}
where $n+m+1\leq5$ and $j\leq3$.
However,
the formulae we have derived are valid 
quite generally; we require only that 
the parameters are nonnegative integers that lead to 
nonnegative exponents.

To perform these integrals, one may use the 
relations
\begin{eqnarray}
2p\cdot k&=&{1\over x-y}(k+xp)^2+{1\over y-x}(k+yp)^2\;\; ,\\
k_\perp^2&=&{z+y\over x-y}(k+xp)^2+{z+x\over y-x}(k+yp)^2\;\; ,\\
k_\perp^2&=&{z+y-x_B\over x-y}(k+xp+q)^2
+{z+x-x_B\over y-x}(k+yp+q)^2\;\; ,
\end{eqnarray}
where $z=k\cdot n$, to systematically cancel
propagators.  With both of the above integrals, one
can continue this process until there are only
three propagators left.  The remaining integral 
can be easily done by contour, leaving only one
nontrivial one-dimensional integral.  
With a suitable change of variables, 
this last integral can always be performed 
with repeated applications of
\begin{equation}
\int_0^1dx\,x^{\alpha-1}(1-x)^{\beta-1}={\Gamma(\alpha)\Gamma(\beta)
\over \Gamma(\alpha+\beta)}\;\; .
\end{equation}

In order to express the value of these integrals,
we define the functions
\begin{eqnarray}
g_n(x,y,\lbrace a_i\rbrace)&=&P^n_n(x_B-x+a_i)\left\lbrack
{2\over\epsilon}\log\left(1-{y\over x_B}\right)-{1\over2}
\log^2\left(1-{y\over x_B}\right)\right\rbrack\nonumber\\
&&+\left\lbrack {2\over\epsilon}-
\log\left(1-{y\over x_B}\right)\right\rbrack
\sum_{i=1}^n{(-1)^i\over i}\\
&&\qquad\times\left\lbrack
1+{\epsilon\over2}\left(S(i)+{1\over i}\right)\right\rbrack
P^n_{n-i}(x_B-x+a_j)\,(x_B-y)^i\;\; ;\nonumber\\
f_n(x,y,\lbrace a_i\rbrace)&=&{1\over y}\left\lbrack
g_n(x,x,\lbrace a_i\rbrace)-g_n(x,x-y,\lbrace a_i\rbrace)\right\rbrack
\;\; ,\\
h_n(x,y,\lbrace a_i\rbrace)&=&
{2\nu\over y}\left\lbrack {1\over x_B-x}g_{n+1}(x,x,\lbrace a_i\rbrace,
x-x_B)\right.\nonumber\\
&&\phantom{{2\nu\over y}\lbrack}\left.
-{1\over x_B-(x-y)}g_{n+1}(x,x-y,\lbrace 
a_i\rbrace,x-x_B)\right\rbrack\;\; ,
\end{eqnarray}
where $i$ runs from 1 to $n$ and 
$P_i^n(x_j)$ is the symmetric product of
order $i$ of the $n$ objects $\lbrace x_j\rbrace$, 
i.e.
\begin{equation}
P_3^4(x_i)=x_1x_2x_3+x_2x_3x_4+x_3x_4x_1+x_4x_1x_2\;\; .
\end{equation}
Note that $g_n$ is a symmetric function of its
last $n$ arguments and 
\begin{equation}
g_n(x,y,\lbrace a_i\rbrace)={\partial\over\partial a_{n+1}}\,g_{n+1}
(x,y,\lbrace a_i\rbrace,a_{n+1})\;\; .
\end{equation}
With these definitions, (\ref{int1}) becomes
\begin{equation}
-{i\over 16\pi^2}\left({Q^2e^\gamma\over4\pi\mu^2}\right)^{-\epsilon/2}
{1\over 2\nu}\sum_{k=1}^m\sum_{\ell=1}^n
{f_{i+j+m-1}(x_k,y_\ell,\lbrace y_\ell\rbrace,\lbrace x_k-x_B\rbrace,
\lbrace a_s\rbrace)
\over \prod_{k'}(x_{k'}-x_k)\prod_{\ell'}(y_{\ell'}-y_\ell)}
\;\; ,
\end{equation}
where the products are defined to exclude the 
singular point, $\lbrace y_\ell\rbrace$ represents
$i$ $y_\ell$'s, $\lbrace x_k\rbrace$ represents 
$m-1$ $x_k$'s, and $\lbrace a_s\rbrace$ represents
all $j$ $a$'s.  The integral in (\ref{int2})
is identical to this result, with $h_{i+j+m-1}$
replacing $f_{i+j+m-1}$.  

With the help of these formulae, the calculation
presented in this paper can be done almost exclusively
on the computer.
\end{appendix}

\end{document}